\documentclass[number,review]{elsarticle}

\pdfoutput=1

\usepackage{amsmath}
\usepackage{graphicx}
\usepackage[colorlinks=false,citecolor=red,linkcolor=black]{hyperref}

\newcommand{\pdx}[1]{\frac{\partial {#1}}{\partial x}}
\newcommand{\pdt}[1]{\frac{\partial {#1}}{\partial t}}
\newcommand{\bb}[1]{\mathrm{\textbf{#1}}}
\newcommand{\bs}[1]{\boldsymbol{#1}}
\newcommand{\abs}[1]{\lvert {#1} \rvert}

\journal{}

\begin{document}
	
	\begin{frontmatter}
		
		
		\title{Re-evaluating efficiency of first-order numerical schemes for two-layer shallow water systems by considering different eigenvalue solutions}

		\author[gradri]{Nino Krvavica\corref{mycorrespondingauthor}}
		\cortext[mycorrespondingauthor]{Corresponding author}
		\ead{nino.krvavica@uniri.hr}

		\address[gradri]{University of Rijeka, Faculty of Civil Engineering, Radmile Matejcic 3, 51000 Rijeka, Croatia}

		\begin{abstract}
			
			The efficiency of several first-order numerical schemes for two-layer shallow water equations (SWE) are evaluated here by considering different eigenvalue solutions. This study is a continuation of our previous work \citep{krvavica2018analytical} in which we have proposed an efficient implementation of a Roe solver for two-layer SWE based on analytical expressions for eigenvalues and eigenvectors. In this work, the accuracy and computational cost of numerical, analytical, and approximated eigenvalue solvers are compared when implemented in Roe, Intermediate Field CaPturing (IFCP) and Polynomial Viscosity Matrix (PVM) schemes. Several numerical tests are performed to examine the overall efficiency of numerical schemes with different eigenvalue solvers when computing two-layer shallow-water flows. The results confirm that analytical eigenvalue solutions are much faster than numerical solvers, with a computational cost closer to approximate expressions. Consequently, the Roe scheme with analytical solutions to the eigenstructure is equally efficient as the IFCP scheme. On the other hand, IFCP and PVM schemes with analytical solutions to eigenvalues are found to be equally efficient as those with approximated expressions. Analytical eigenvalues show slightly better results when dealing with larger density differences between the layers.

		\end{abstract}
		
		
	\end{frontmatter}

\section{Introduction}

Two-layer shallow-water equations (SWE) are widely used to simulate geophysical flows in stratified conditions. Some examples of a two-layer configuration include exchange flows in sea straits \citep{castro2001q,castro2004numerical}, highly stratified estuaries \citep{krvavica2017numerical,krvavica2016salt}, as well as various types of gravity currents \citep{la2012two,adduce2011gravity}, such as mudflows \citep{canestrelli2012mass}, debris flows \citep{pelanti2008roe,majd2014lhllc}, submarine avalanches \citep{fernandez2008new,luca2009two,pitman2005two}, and pyroclastic flows \citep{doyle2011two}. Although such processes can be described more accurately by 3D Navier-Stokes equations, two-layer shallow water models make a popular alternative because of their simplicity and a significantly lower computational cost.

Two-layer SWEs are mathematically defined as hyperbolic systems of coupled conservation laws with source terms, or so-called balance laws \citep{castro2001q}. These equations are challenging to solve numerically because of the layer coupling and non-conservative source terms accounting for the variable geometry, friction, or entrainment. Over the last two decades, a numerical resolution of two-layer SWE has been an object of intense research \citep{castro2001q,castro2004numerical,kurganov2009central,castro2010some,murillo2010exner,bouchut2010robust,fernandez2011intermediate,lee2011fast,canestrelli2012restoration,chiapolino2018models}.
A popular choice for numerical resolution of two-layer SWEs are finite volume methods (FVM), and among them a family of path-conserving schemes \citep{toumi1992weak,bermudez1994upwind,castro2001q,pares2004well,pares2006numerical}. 

Implementation of the path-conservative schemes involves a numerical viscosity matrix, which is usually derived from some or all eigenvalues of a corresponding Jacobian of the flux matrix. The choice of the numerical viscosity matrix determines the numerical diffusion and accuracy of the scheme.
Since analytical expressions for eigenvalues of two-layer SWE systems were considered unavailable until recently \cite{castro2004numerical,pelanti2008roe,abgrall2009two,fernandez2011intermediate,sarno2017some}, either approximate expressions \cite{schijf1953theoretical,abgrall2009two} or numerical algorithms were used instead. Unfortunately, numerical algorithms, such as root-finding and eigensolver methods, make schemes computationally (too) demanding. For example, several studies evaluating the efficiency of first-order schemes in solving two-layer flows \cite{castro2010some,castro2012class,fernandez2011intermediate}, find Roe schemes, which are based on all eigenvalues, to be the most accurate, but far less efficient then some other first-order schemes, mainly because of excessive computational costs when performing a full spectral decomposition by a numerical eigensolver.

Recently, a new solution \cite{krvavica2018analytical} to the efficiency problem for the Roe scheme was proposed. This new approach introduced a semi-analytical implementation of the Roe scheme based on simple closed-form solutions for the eigenvalues and eigenvectors \cite{krvavica2018analytical}. New scheme, named A-Roe, was found to be much faster than the numerical implementation of the Roe scheme, while producing equally accurate results \cite{krvavica2018analytical}. An additional advantage is that closed-form solutions enable a direct and accurate prediction of complex eigenvalues and implementation of a corrective algorithm for the loss of hyperbolicity \cite{krvavica2018analytical}.

Our previous paper \cite{krvavica2018analytical} evaluated the performance of the A-Roe scheme based on analytical expressions for eigenvalues and eigenvectors, and compared its efficiency to Lax-Friedrichs (LF) and GFORCE scheme, as well as the Polynomial Viscosity Matrix redefinition of the Roe scheme (PVM-Roe) \cite{castro2012class} and Intermediate Field CaPturing scheme (IFCP) \cite{fernandez2011intermediate}. The former two do not use any eigenvalue information, whereas the latter two use all four eigenvalues \cite{castro2010some,castro2012class, fernandez2011intermediate}. In the case of PVM-Roe and IFCP schemes, the classical implementation based on approximated eigenvalues was considered.  Furthermore, our previous paper \cite{krvavica2018analytical} focused on overall advantages of using analytical eigenvalues; primarily, increased computational speed and hyperbolicity correction algorithm.

The present study is a natural continuation of the previous paper \cite{krvavica2018analytical}. One evident question emerged from the main conclusions of the previous paper - if analytical expressions for the eigenstructure significantly improved the efficiency of the Roe scheme, what effect will they have on the performance of other first-order schemes that are also based on some or all eigenvalues? Clearly, the efficiency of those numerical schemes should be re-evaluated. 

The main aim of this study is to investigated the sensitivity of several other numerical schemes, namely PVM-2U \cite{castro2012class} and IFCP \cite{fernandez2011intermediate}, on the choice of eigenvalues and to evaluate potential benefits of using analytical eigenvalues instead of recommended approximated expressions. 
For this purpose, the accuracy and computational speed of recently proposed closed-form eigenvalue solutions are carefully compared against two available alternatives - the numerical eigensolvers and approximated expressions for eigenvalues. Next, the sensitivity of numerical schemes to the choice of an eigenvalue solver is assessed. And finally, the overall computational efficiency of Roe, IFCP, and PVM-2U schemes with different eigenvalues is evaluated by performing several numerical tests which consider different density differences between the layers and different channel geometries.
Some significant remarks on the implementation of numerical schemes, which increase their efficiency, are also presented.

\section{Two layer shallow-water flow: Theory, eigenvalues, and numerical schemes}

\subsection{Governing system of equations}

A one-dimensional (1D) two-layer shallow-water flow in prismatic channels with rectangular cross-sections of constant width is considered for all tests. The governing system of equations written in a general vector form is repeated here in a more compact form for context and reproducibility \citep{castro2001q}:
\begin{equation}
\pdt{\bb{w}} + \pdx{\bb{f}(\bb{w})} = 
\bb{B}(\bb{w})\pdx{\bb{w}} 
+ \bb{g}(\bb{w}),
\label{eq:consys1}
\end{equation}
where $x$ refers to the axis of the channel and $t$ is time. The vector of conserved quantities $\bb{w}$ is defined as:
\begin{equation}
\bb{w} = 
\begin{Bmatrix}
h_{1} \quad   q_{1} \quad  h_{2} \quad  q_{2}
\end{Bmatrix}^T,
\quad 
\end{equation}
where $h_j$ is the layer thickness (or depth), $q_j=h_j u_j$ is the layer flow rate per unit width, and index $j=1,2$ denotes the respective upper and lower layer. The flux vector $\bb{f(w)}$ is:
\begin{equation}
\bb{f}(\bb{w}) = 
\begin{Bmatrix}
q_{1} \quad
\tfrac{q^{2}_{1}}{h_{1}} + \tfrac{g}{2}h^{2}_{1}  \quad
q_{2} \quad
\tfrac{q^{2}_{2}}{h_{2}} + \tfrac{g}{2}h^{2}_{2}
\end{Bmatrix}^T,
\end{equation}
where $g$ is acceleration of gravity. Matrix $\bb{B(w)}$ is a result of coupling the two-layer system, defined as \citep{castro2001q}:
\begin{equation}
\bb{B}(\bb{w}) = 
\begin{bmatrix}
0 & 0 & 0 & 0 \\
0 & 0 & -c_1^2 & 0 \\
0 & 0 & 0 & 0 \\
-rc_2^2 & 0 & 0 & 0
\end{bmatrix},
\end{equation}
where $r=\rho_1/\rho_2 < 1$ is the ratio between the upper layer density $\rho_1$ and the lower layer density $\rho_2$, and $c_j^2=gh_j$ is propagation celerity of internal and external perturbations (waves), for $j=1,2$.
Finally, the bathymetry source term $\bb{g(w)}$ is defined as follows \citep{castro2001q}:
\begin{equation}
\bb{g}(\bb{w}) = 
\begin{Bmatrix}
0 \quad
-gh_1\frac{\textrm{d}b}{\textrm{d}x} \quad
0 \quad
-gh_2\frac{\textrm{d}b}{\textrm{d}x}
\end{Bmatrix}^T,
\end{equation}
where $b$ is the bed elevation.

\subsection{Eigenvalues}

The system given by Eq.~(\ref{eq:consys1}) can be rewritten in the following quasi-linear form \cite{castro2001q}:
\begin{equation}
\frac{\partial \mathbf{w}}{\partial t} + \boldsymbol{\mathcal{A}}(\mathbf{w})\frac{\partial \mathbf{w}}{\partial x} = 
\mathbf{g}(\mathbf{w}),
\label{eq:consys2}
\end{equation}
where 
\begin{equation}
\boldsymbol{\mathcal{A}(\mathbf{w}}) = 
\frac{\partial \mathbf{f(w)}}{\partial \mathbf{w}} - \mathbf{B}(\mathbf{w}) = 
\mathbf{J}(\mathbf{w}) - \mathbf{B}(\mathbf{w})
\label{eq:AJB}
\end{equation}
is the pseudo-Jacobian matrix that contains the flux gradient terms as well as the coupling terms:
\begin{equation}
\boldsymbol{\mathcal{A}(\mathbf{w})} =
\begin{bmatrix}
0 				& 1 	& 0 			& 0 \\
c_1^2 - u_1^2 	& 2u_1 	& c_1^2 		& 0 \\
0 				& 0 	& 0 			& 1 \\
rc_2^2 			& 0 	& c_2^2 - u_2^2	& 2u_2.
\end{bmatrix}
\label{eq:A}
\end{equation}

The four eigenvalues of $\boldsymbol{\mathcal{A}(\mathbf{w})}$ define the propagation speeds of barotropic (external) and baroclinic (internal) perturbations.
In most geophysical flows, one of two external eigenvalues is negative $\lambda_{ext}^{-} < 0$, while the other is positive $\lambda_{ext}^{+} > 0$ \cite{abgrall2009two}.
Eigenvalues can be computed using numerical solvers, approximated expressions or analytical solutions.

\subsubsection{Numerical eigenvalues}

Eigenvalues and eigenvectors of $\boldsymbol{\mathcal{A}(\mathbf{w})}$ can be numerically obtained by solving the following equation:
\begin{equation}
\boldsymbol{\mathcal{A}} \bb{K} = 
\bb{K}
\boldsymbol{\Lambda}.
\label{eq:eigendecomposition}
\end{equation}
where $ \boldsymbol{\Lambda}$ is a $4\times 4$ diagonal matrix whose coefficient are the eigenvalues $\lambda_k, k=1,..,4$, and $\bb{K}$ is matrix whose columns are the corresponding right eigenvectors. Usually a QR algorithm is used for this purpose \cite{lapack}.
 
\subsubsection{Approximated eigenvalues}

The following approximation derived under the assumption of $r\approx 1$ and $u_1\approx u_2$ are usually used for computing the internal and external eigenvalues \citep{schijf1953theoretical}:
\begin{equation}
\lambda_{ext}^{\pm} = U_1 \pm \sqrt{g(h_1+h_2)}
\label{eq:eig_ext}
\end{equation}
\begin{equation}
\lambda_{int}^{\pm} = U_2 \pm \sqrt{g(1-r) \frac{h_1 h_2}{h_1 + h_2} \left[ 1 - \frac{(u_1 - u_2)^2}{g(1-r)(h_1 + h_2)}\right]},
\label{eq:eig_int}
\end{equation}
with
\begin{equation}
U_1 = \frac{h_1 u_1 + h_2 u_2}{h_1 + h_2} \quad \textrm{and} \quad
U_2 = \frac{h_1 u_2 + h_2 u_1}{h_1 + h_2}.
\end{equation}

Note that Eqs.~(\ref{eq:eig_ext}) and (\ref{eq:eig_int}) are valid only when dealing with two layers of similar densities ($r = \rho_1/\rho_2 \approx 1$) and when velocities in both layers are comparable ($u_1\approx u_2$). Those conditions are found in some stratified flows in nature, such as exchange flows through sea straits \citep{castro2004numerical,chakir2009roe} or some cases of highly stratified estuaries \cite{krvavica2016salt}. However, for geophysical flows characterized by a larger relative density difference, such as granular, debris or mud flows, the approximated values may significantly deviate from exact values and cannot accurately predict a possible hyperbolicity loss \cite{abgrall2009two,sarno2017some,krvavica2018analytical}.

\subsubsection{Analytical eigenvalues}

Recently, a simple closed-form approach for computing real roots of the characteristic quartic Eq.~(\ref{eq:polynomial}) was proposed \cite{krvavica2018analytical}. The solutions are based on Ferrari's formulas \citep{abramowitz1972}, they consist of eight simple evaluations, and are repeated here for consistency and clarity. A detailed derivation of these equations is available in \cite{krvavica2018analytical}. 
The proposed closed-form solutions to eigenvalues are given in terms of coefficients $a, b, c$ and $d$ of a characteristic polynomial of matrix $\boldsymbol{\mathcal{A}(\mathbf{w})}$ \cite{krvavica2018analytical}:
\begin{equation}
p(\lambda) = \lambda^4 + a\lambda^3 + b\lambda^2 + c\lambda + d
\label{eq:polynomial}
\end{equation} 
with:
\begin{equation}
a = -2 \left( u_1 + u_2 \right), \\
\label{eq:a}
\end{equation}
\begin{equation}
b = u_1^2  - c_1^2 + 4u_1u_2 +  u_2^2 - c_2^2,
\label{eq:b}
\end{equation}
\begin{equation}
c = - 2u_2 \left(u_1^2 - c_1^2 \right) - 2u_1 \left( u_2^2 - c_2^2 \right) ,
\label{eq:c}
\end{equation}
\begin{equation}
d = \left( u_1^2 - c_1^2 \right) \left( u_2^2 - c_2^2 \right) - rc_1^2c_2^2 .
\label{eq:d}
\end{equation}

Real eigenvalues are then computed by the following expressions \cite{krvavica2018analytical}:
\begin{equation}
\lambda_{ext}^{\pm} = \lambda_{4,1} = - \frac{a}{4} \pm \frac{ \sqrt{Z} + \sqrt{- A - Z \mp \frac{B}{\sqrt{Z}} } }{2} ,
\label{eq:solution1}
\end{equation}
\begin{equation}
\lambda_{int}^{\pm} = \lambda_{3,2} = - \frac{a}{4} \pm \frac{ \sqrt{Z} - \sqrt{- A - Z \mp \frac{B}{\sqrt{Z}} } }{2} .
\label{eq:solution2}
\end{equation}
where
\begin{equation}
Z =  \frac{1}{3} \left( 2 \sqrt{\Delta_0} \cos \frac{\phi}{3}  - A \right),
\label{eq:Zcoeff}
\end{equation}
\begin{equation}
\phi = \arccos \left( \frac{\Delta_1}{2 \Delta_0 \sqrt{\Delta_0}}\right),
\label{eq:S_cubic}
\end{equation}
with
\begin{equation}
A = 2b - \frac{3a^2}{4},
\label{eq:Acoeff}
\end{equation}
\begin{equation}
B = 2c - ab + \frac{a^3}{4} .
\label{eq:Bcoeff}
\end{equation}
and
\begin{equation}
\Delta_0 = b^2 + 12d - 3ac, 
\label{eq:D0}
\end{equation}
\begin{equation}
\Delta_1 = 27a^2d - 9abc + 2b^3 - 72bd + 27c^2.
\label{eq:D1}
\end{equation}

Note, that it is possible to combine these equations into a single explicit solution in terms of conserved variables, but the resulting formula would certainly be too extensive to be presented in a journal format, and probably not optimized to be implemented in a computational algorithm. However, an example of such expanded formulation is available \cite{wikipedia}.

\subsubsection{Nature of eigenvalues}

Since the relative density difference $r$ has a major influence on internal eigenvalues, they are usually smaller than the external ones. Therefore, the following indexing and order of eigenvalues will be used herein:
\begin{equation}
\lambda_1 = \lambda_{ext}^{-}, \quad \lambda_2 = \lambda_{int}^{-}, \quad
\lambda_3 = \lambda_{int}^{+}, \quad \lambda_4 = \lambda_{ext}^{+}
\end{equation}

It is worth mentioning that the external eigenvalues $\lambda_{ext}^{\pm}$ are always real \cite{ovsyannikov1979two,abgrall2009two}; however, at sufficiently large relative velocities $\Delta u = \lvert u_1 - u_2 \rvert$, as well as for very low or very high relative densities $r$, the internal eigenvalues $\lambda_{int}^{\pm}$ may become complex and the governing system loses its hyperbolic character \cite{ovsyannikov1979two,castro2011numerical}. 

Although analytical expressions given by Eqs. (\ref{eq:solution1}) and (\ref{eq:solution2}) are valid only for real roots, a simple hyperbolicity correction algorithm can be incorporated, which verifies whether the term in Eq.~(\ref{eq:S_cubic}) satisfies the condition $\abs{\frac{\Delta_1}{2 \Delta_0 \sqrt{\Delta_0}}} < 1$, and if not, iteratively corrects the velocities until a hyperbolic state is recovered. This algorithm is not a subject of the current study, but its description and details are available in \cite{krvavica2018analytical}.

\subsection{Numerical schemes}

A class of path-conservative schemes are considered here to approximate the governing equations for two-layer shallow water flow \citep{pares2006numerical}. A first order accurate path-conservative scheme for Eq.~(\ref{eq:consys2}) may be written as follows \citep{pares2006numerical}:
\begin{equation}
\bb{w}_i^{n+1} = \bb{w}_i^n - \frac{\Delta t}{\Delta x} \left(\bb{D}_{i-1/2}^{+} + \bb{D}_{i+1/2}^{-} \right)
\label{eq:pathcons}
\end{equation}
where $\Delta x$ and $\Delta t$ are the respective spatial and time increment (considered constant here for simplicity), $\bb{w}_i^n$ denotes the approximate cell-averaged values of the exact solution obtained by the numerical scheme at cell $I_i = [x_{i-1/2}, x_{i+1/2}]$ in time $t^n = n\Delta t$, and matrices $\bb{D}_{i+1/2}^{\pm}$ are continuous functions of conserved variables $\bb{D}^{\pm} \left( \bb{w}_i^n, \bb{w}_{i+1}^n \right)$.

For the governing system of equations, a generalized numerical scheme based on Roe linearisation  \cite{toumi1992weak} may be written by Eq.~(\ref{eq:pathcons}), with:
\begin{equation}
\begin{aligned}
\bb{D}_{i+1/2}^{\pm} =& \dfrac{1}{2} \Big[ \bb{f}(\bb{w}_{i+1}^n) - \bb{f}(\bb{w}_i^n) + \bb{B}_{i+1/2}(\bb{w}_{i+1}^n - \bb{w}_i^n) - \bb{g}_{i+1/2} \\
& \pm \bb{Q}_{i+1/2} \left( \bb{w}_{i+1}^n - \bb{w}_i^n - \bs{\mathcal{A}}_{i+1/2}^{-1} \bb{g}_{i+1/2}\right) \Big]
\end{aligned}
\label{eq:any_scheme}
\end{equation}
where $\boldsymbol{Q}_{i+1/2}$ is a numerical viscosity matrix that determines the numerical diffusivity of the results, and whose choice depends on a particular scheme \cite{fernandez2011intermediate}. Matrices and vectors $\bb{B}_{i+1/2}$, $\bb{A}_{i+1/2}$, and $\bb{g}_{i+1/2}$ correspond to $\bb{B}(\bb{w}_{i+1}^n, \bb{w}_i^n)$, $\bb{A}(\bb{w}_{i+1}^n, \bb{w}_i^n)$, and $\bb{g}(\bb{w}_{i+1}^n, \bb{w}_i^n)$, respectively, evaluated at the cell interface after a suitable Roe linearization is performed (see \cite{castro2001q} and \cite{krvavica2018analytical} for details). 

The viscosity matrix in Roe schemes coincides with the absolute pseudo-Jacobian matrix \citep{castro2001q}:
\begin{equation}
\boldsymbol{Q}_{i+1/2} = \lvert \boldsymbol{\mathcal{A}}_{i+1/2} \rvert
\label{eq:Q_Roe}
\end{equation}
and the absolute value of $\boldsymbol{\mathcal{A}}_{i+1/2}$ can be directly obtained from:
\begin{equation}
\lvert \boldsymbol{\mathcal{A}}_{i+1/2} \rvert = 
\bb{K}_{i+1/2}
\lvert \boldsymbol{\Lambda}_{i+1/2} \rvert
\bb{K}_{i+1/2}^{-1}.
\label{eq:A_abs}
\end{equation}
where $\lvert \boldsymbol{\Lambda}_{i+1/2} \rvert$ is a $4\times 4$ diagonal matrix whose coefficient are the absolute eigenvalues $\lvert \lambda_k \rvert, k=1,..,4$.

When analytical eigenvalues are unavailable, Roe schemes require either approximation of eigenvalues and corresponding eigenvectors or a numerical decomposition to obtain the eigenstructure of matrix $\boldsymbol{\mathcal{A}}_{i+1/2}$. The latter is computational expensive, while the former is less accurate. In both cases, an additional re-composition of the viscosity matrix from eigenstructure is required (see Eq.~(\ref{eq:A_abs})), which imposes an additional computational cost. This drawback has motivated a development of numerical schemes in which the viscosity matrix $\boldsymbol{Q}_{i+1/2} $ is directly approximated from pseudo-Jacobian matrix. 

One possible way to construct such a scheme is to approximate the viscosity matrix by a polynomial function. Those class of methods are called Polynomial Viscosity Matrix (PVM) \cite{castro2012class}.
PVM methods define the viscosity matrix through a general polynomial evaluation of the pseudo-Jacobian matrix, given by:
\begin{equation}
\boldsymbol{Q}_{i+1/2} = p_l \left( \mathcal{A}_{i+1/2} \right)
\end{equation}
where $p_l(x)$ is a polynomial of degree $l$
\begin{equation}
p_l(x) = \sum_{j=0}^{l} \alpha_j x^j
\end{equation}
The main advantage of PVM schemes is that a full spectral decomposition is not required.

In \cite{castro2012class} several different PVM schemes were presented and tested, some of them were derived from other popular Riemann solvers redefined under the PVM formulation, and some of them were newly proposed. In our previous paper \cite{krvavica2018analytical} we chose a PVM redefinition of the Roe scheme (PVM-Roe) that is based on all four eigenvalues, and here we choose a scheme called PVM-2U that is based only on two external eigenvalues. In \cite{castro2012class}, PVM-2U proved to be the most efficient throughout their numerical tests and is, therefore, selected here for further analysis and comparison regarding the effects of the eigenvalue solvers.

The viscosity matrix of PVM-2U may be computed as \cite{castro2012class}:
\begin{equation}
\boldsymbol{Q}_{i+1/2} = \alpha_0 \bb{Id} + \alpha_1 \boldsymbol{\mathcal{A}}_{i+1/2} + \alpha_2 \boldsymbol{\mathcal{A}}_{i+1/2}^2,
\label{eq:PVM2_Q}
\end{equation}
where coefficients $\alpha_k$, $k=0,1,2$ are derived from two external eigenvalues (for more details see \cite{castro2012class}). 

A similar scheme called Intermediate Field Capturing Parabola (IFCP) \cite{fernandez2011intermediate} was also derived from the family of PVM schemes. In contrast to PVM-2U, the IFCP scheme uses both internal and external eigenvalue information, and should be more accurate than PVM-2U with a minor increase in the computational cost. The IFCP scheme is also defined by Eq.~(\ref{eq:PVM2_Q}), where $\alpha_k$, $k=0,1,2$ are derived from two external and one internal eigenvalues (for more details see \cite{fernandez2011intermediate}).

It should be emphasized that all three schemes considered in this study (Roe, PVM-2U, IFCP) are well balanced for water at rest solutions and linearly stable under the CFL condition \cite{castro2001q,castro2012class}:
\begin{equation}
\frac{\Delta t}{\Delta x} \max \abs{\lambda_{j,i+1/2}} = \gamma \leq 1
\end{equation}

Regardless of a particular scheme, the eigenvalues may be computed either by a numerical decomposition, analytical (closed-form) or approximated expressions. In most published studies \cite{fernandez2011intermediate,castro2012class}, approximated expressions are recommended for both schemes from the PVM family when dealing with two-layer SWEs, as a more efficient choice in comparison to numerical eigensolvers. This study examines the benefits of using the analytical solutions instead.

\subsection{Remarks about the implementation of the schemes}
\label{sec:remarks}

There are some modifications of numerical schemes that can be made to optimize the algorithms and improve their computational performance. 

The first optimization deals with the Roe scheme and re-composition of the viscosity matrix given by Eq.~(\ref{eq:A_abs}). 
Although analytical closed-form solutions to $\bb{K}^{-1}$ are available in the Appendix of \cite{krvavica2018analytical}, it is computationally faster to rewrite Eq.~(\ref{eq:A_abs}) as
\begin{equation}
\bb{K}_{i+1/2}^T \abs{\bs{\mathcal{A}}_{i+1/2}}^T  = (\bb{K}_{i+1/2} \vert \boldsymbol{\Lambda}_{i+1/2} \vert)^T ,
\label{eq:Aabs_T}
\end{equation} 
which corresponds to a general matrix equation $\bb{Ax}=\bb{B}$, solve it numerically for $ \abs{\bs{\mathcal{A}}_{i+1/2}}^T$ (for example, by a LAPACK routine \textit{gesv} \citep{lapack}), and then transpose it. This is about 2-3 times faster than finding the inverse of $\bb{K}_{i+1/2}$ and performing matrix multiplication to obtain the viscosity matrix as written in Eq.~(\ref{eq:A_abs}).

A second optimization is available for the family of PVM schemes. Using Eq.~(\ref{eq:AJB}) and the usual Roe linearization of the pseudo-Jacobian matrix:
\begin{equation}
\bb{J}(\bb{w}_{i+1}, \bb{w}_i) \cdot \left(\bb{w}_{i+1} - \bb{w}_i\right) = \bb{f}(\bb{w}_{i+1}) - \bb{f}(\bb{w}_{i})
\end{equation}	
term $\bb{Q}_{i+1/2} \left( \bb{w}_{i+1}^n - \bb{w}_i^n \right)$ given in Eq.~(\ref{eq:any_scheme}) may be replaced by:
\begin{equation}
\bb{C}_{i+1/2} \left[ \bb{f}(\bb{w}_{i+1}^n) - \bb{f}(\bb{w}_i^n) - \bb{B}_{i+1/2}(\bb{w}_{i+1}^n - \bb{w}_i^n) \right] \\
\label{eq:optimal_scheme}
\end{equation}
where $\bb{C}_{i+1/2} = \bb{Q}_{i+1/2} \bs{\mathcal{A}}_{i+1/2}^{-1}$ lowers the order of a viscosity matrix by one, and is already needed for the source term discretization. In other words, full viscosity matrix $\bb{Q}_{i+1/2}$ is not required for the family of PVM schemes; instead, only $\bb{C}_{i+1/2}$ is computed. For both PVM-2U and IFCP, $\bb{C}_{i+1/2}$ is defined as:
\begin{equation}
\bb{C}_{i+1/2} = \alpha_0 \bs{\mathcal{A}}_{i+1/2}^{-1} + \alpha_1 \bb{Id} + \alpha_2 \boldsymbol{\mathcal{A}}_{i+1/2},
\label{eq:PVM2_C}
\end{equation}
In this way, computation of the square of $\bs{\mathcal{A}}_{i+1/2}$ (see Eq.~(\ref{eq:PVM2_Q})) is avoided.

\section{Results}

Several numerical tests are presented to evaluate the efficiency of Roe, IFCP, and PVM-2U schemes with different eigenvalue solvers. First, the accuracy and computational speed of numerical, analytical and approximated eigenvalue solvers is analysed. Next, the performance of numerical schemes in computing the numerical viscosity matrix is examined, as well as their sensitivity to the choice of eigenvalues. Finally, five numerical tests are given to analyse the overall efficiency of different numerical schemes in computing two-layer shallow-water flows.

In particular, three eigensolver algorithms are examined:
\begin{itemize}
	\item N-Eig uses a numerical eigenvalue solver which decomposes a general square matrix into a diagonal matrix $\bs{\Lambda}$ whose elements are eigenvalues, and matrix $\bb{K}$ whose columns are right eigenvectors. 
	This algorithm is implemented in Python using Numpy function \texttt{numpy.linalg.eig} which is based on the \texttt{geev} LAPACK routines written in FORTRAN \cite{lapack}. 
	\item A-Eig is an analytical eigenvalue solver based on a closed-form solution to the roots of the characteristic quartic given by Eqs.~(\ref{eq:solution1}) and (\ref{eq:solution2})
	\item E-Eig only estimates eigenvalues based on the approximations given by Eqs.~(\ref{eq:eig_ext}) and (\ref{eq:eig_int})
\end{itemize}

By combining different eigensolvers with Roe, IFCP, and PVM-2U schemes, the following numerical algorithms for computing the viscosity matrix and solving two-layer SWEs are chosen for the efficiency analysis: N-Roe, A-Roe, E-Roe, A-IFCP, E-IFCP, A-PVM2, and E-PVM2 schemes. 

The accuracy of algorithms and schemes are evaluated by using either absolute error AE, relative bias error RBE,  or root relative square error RRSE, defined respectively as:
\begin{equation}
AE_{\Phi} = \lvert \Phi - \Phi^{ref} \rvert
\label{eq:AE}
\end{equation}
\begin{equation}
RBE_{\Phi} = \frac{\Phi - \Phi^{ref}}{\Phi^{ref}}
\label{eq:BE}
\end{equation}
\begin{equation}
RRSE_{\Phi} = \frac{\sqrt{\sum_{n=1}^{M} \left[ \Phi(x_n, t_{end}) - \Phi^{ref}(x_n, t_{end}) \right]^2}}{\sqrt{\sum_{n=1}^{M} \Phi^{ref}(x_n, t_{end})^2}},
\label{eq:RRSE}
\end{equation}
where $\Phi$ is the evaluated parameter (e.g. eigenvalue $\lambda$, numerical flux $f$, depth $h$ or velocity $u$), $\Phi^{ref}$ are the corresponding reference values, and $M$ is a number of spatial points. 

All numerical algorithms have been implemented in Python 3.6 and vectorized using the Numpy package. The tests have been performed on 64-bit Windows 10 machine with Intel Core i7-3770 3.4 GHz processor. All algorithms for computing the eigenstructure and viscosity matrix are freely available on Github \cite{krvavica2019github}.

\subsection{Computing eigenvalues: Accuracy and computational speed}

This test examines the accuracy and computational speed (CPU time) of corresponding three different algorithms for computing eigenvalues N-Eig, A-Eig, and E-Eig. Although eigenvalues computed by the numerical eigensolver are not \textit{exact}, their errors are of the order of round-off errors. Therefore, the numerical results are used as a reference when evaluating the accuracy of analytical and approximated eigenvalues. 

Since the main idea is to evaluate algorithms for solving eigenvalues as a integral part of Roe, IFCP and PVM-2U schemes for two-layer SWEs, physically realistic flow parameters (which always produce real eigenvalues) are chosen for this test. Therefore, a large set of parameters ($N=10^6$) is randomly generated from a given range: $ 1.0 < h_{1,2} < 2.0$ m and $ -0.3 < u_{1,2} < 0.3$ m s$^{-1}$. Different density ratios are selected, namely $r = 0.98, 0.9, 0.7, 0.5,$ and $0.3$, with $g$ set to $9.8$ m s$^{-2}$.

\begin{figure}[htbp]
	\center
	\includegraphics[width=6cm]{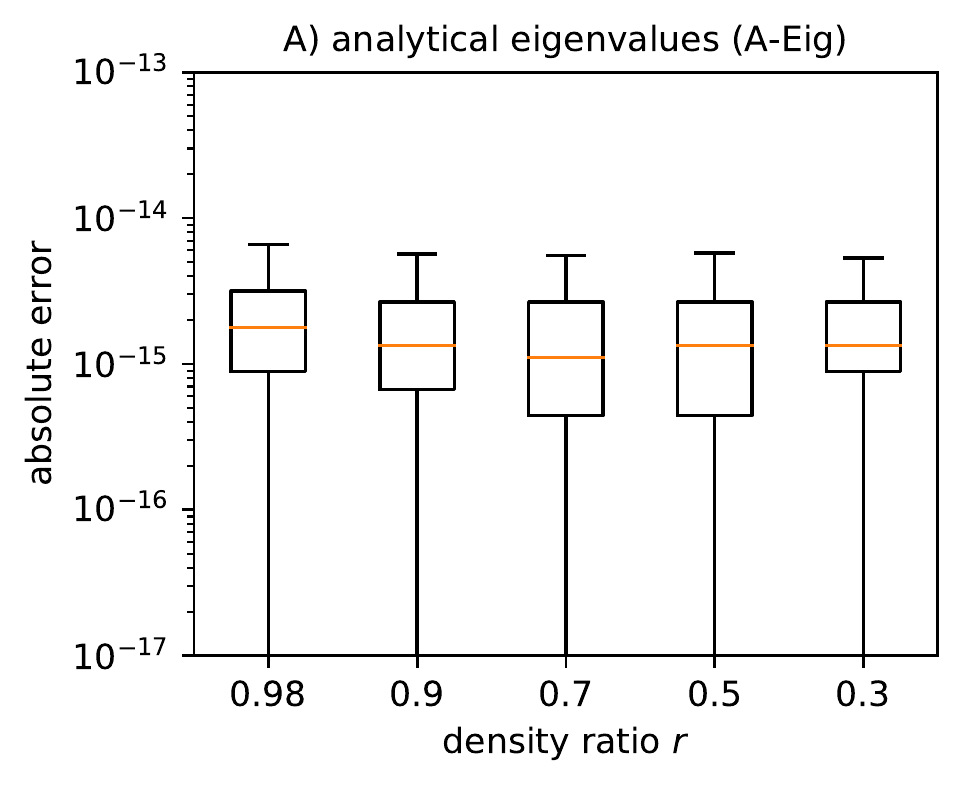}
	\hfill	
	\includegraphics[width=6cm]{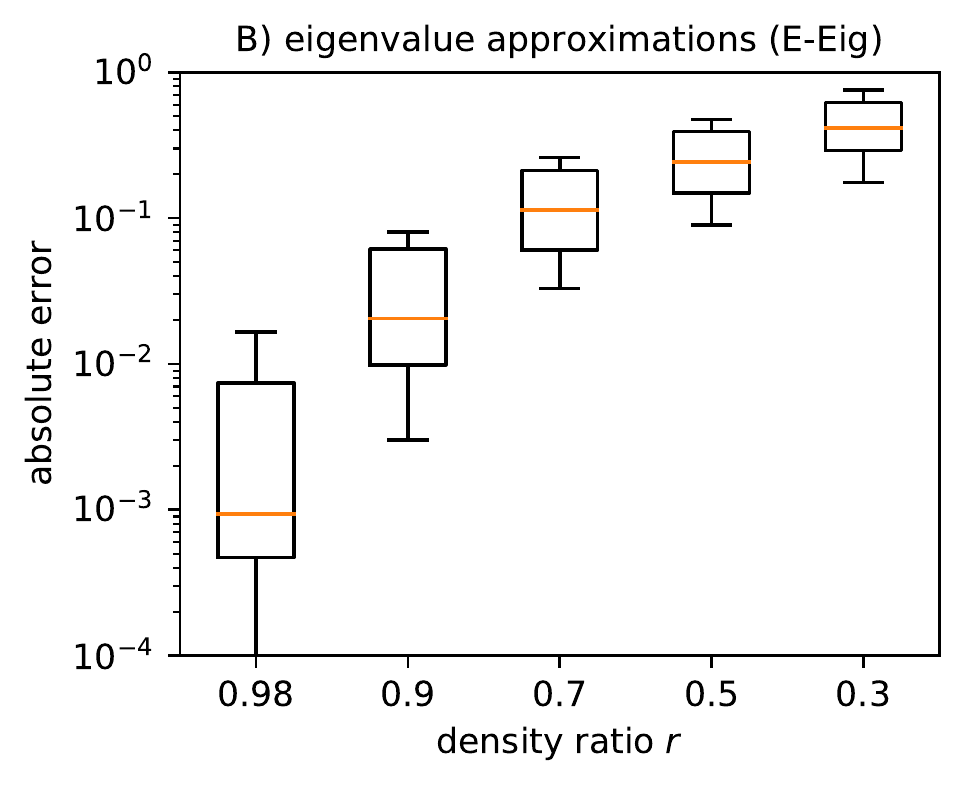}
	\caption{Boxplot of the absolute errors when computing eigenvalues by: A) the analytical closed-form solutions (A-Eig) and B) eigenvalue approximations (E-Eig). Boxes denote the interquartile range and median value, while whiskers denote min and max values.}
	\label{fig:eig_errors}
\end{figure}

Fig.~\ref{fig:eig_errors} illustrates the statistical representation of the absolute errors of A-Eig and E-Eig algorithms computed by Eq.~(\ref{eq:AE}) for one million sets of independent flow parameters. The average error of the analytic solver is around $10^{-15}$ and the maximum errors are below $10^{-14}$ which is close to a round-off error of the numerical eigensolver. In other words, both these solvers produce almost identical results. The average errors of the approximated expressions are more noticeable, and they increase with density difference. For $r=0.98$, average errors are around $10^{-3}$, whereas for $r=0.3$, average errors are larger than $10^{-1}$. This is expected, since eigenvalue approximations are derived under the assumption of $r\approx1$ and $u_1 \approx u_2$, and become less accurate as parameters deviate from these assumptions.

\begin{figure}[htbp]
	\center
	\includegraphics[width=6cm]{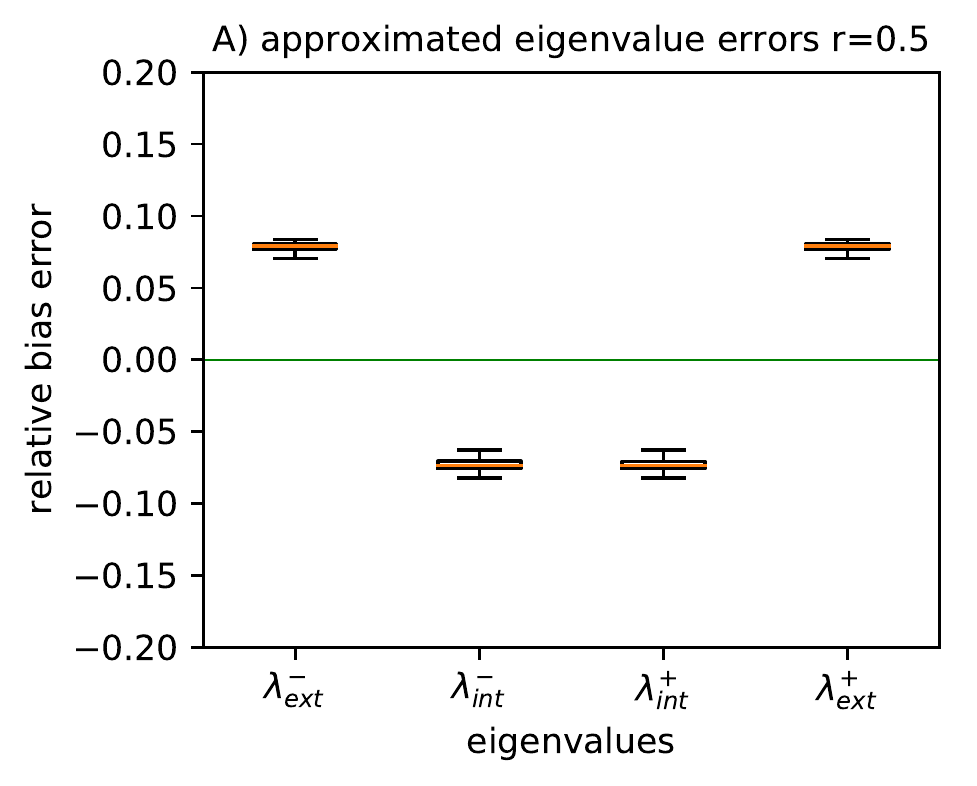}
	\hfill	
	\includegraphics[width=6cm]{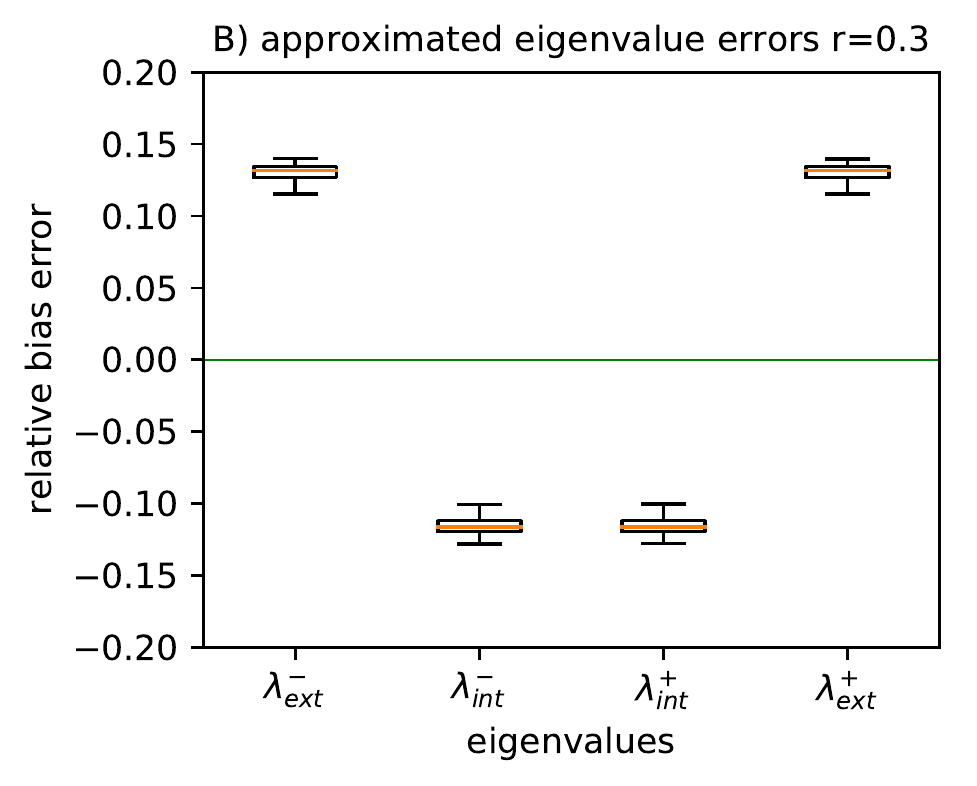}
	\caption{Boxplot of the relative bias errors when using approximated eigenvalues E-Eig for: A) $r=0.5$ and B) $r=0.3$. Boxes denote the interquartile range and median value, while whiskers denote min and max values.}
	\label{fig:eig_bias}
\end{figure}

To examine the lack of accuracy of the eigenvalues approximations, Fig.~\ref{fig:eig_bias} illustrates the relative bias errors of individual eigenvalues for small $r$. It seems that the eigenvalue approximations always overestimate external eigenvalues, and underestimate the internal ones. Contrary to approximated expressions given by Eq.~\ref{eq:eig_ext} that suggest how $r$ only affects the internal eigenvalues, it is clear from this analysis that $r$ equally affects external and internal eigenvalues.

Table~\ref{tab:eig_cpu} shows the computational speed of three eigenvalue algorithms. The fastest algorithm is the E-Eig (based on approximated eigenvalues) that needed 0.11 s, followed by the analytical solver A-Eig with 0.29 s, and finally the numerical solver N-Eig with 4.14 s. Notice that both approximated and closed-form eigenvalue solvers are one order of magnitude faster than the numerical solver. These results are in agreement with \cite{krvavica2018analytical} and suggest that the prevailing opinion in the scientific community about the "computational complexity" of the analytical solver is not justified. Although A-Eig needs double the time of E-Eig, it is considerably more accurate, as shown in Fig.~\ref{fig:eig_errors}.

\begin{table}[htbp]
	\small
	\centering
	\caption{CPU times when computing one million sets of eigenvalues by numerical N-Eig, analytical A-Eig, and approximated E-Eig eigenvalue solvers (best of 5 runs).}
	\begin{tabular}{clll}
		\hline
		Eigensolver & N-Eig   & A-Eig & E-Eig \\
		\hline
		CPU time (ms)  & 4136.5  & 286.8     & 109.7 \\
		\hline
	\end{tabular}%
	\label{tab:eig_cpu}%
\end{table}%

\subsection{Computing numerical viscosity matrix: Accuracy and computational speed}

The next test examines the accuracy and the computational speed of different implementations of Roe, IFCP, and PVM-2U schemes. The implementation differs in the choice of a particular eigenvalue algorithm, presented in the previous subsection. The main goal is to investigate how do eigenvalue algorithms affect the efficiency of Roe, IFCP, and PVM-2U schemes when computing the corresponding numerical viscosity matrix, and to analyse their sensitivity to a particular choice of the eigenvalue solver.

Same as in the previous example, physically realistic flow parameters are chosen. One million sets of parameters are randomly generated from a given range: $ 1.0 < h_{1,2} < 2.0$ m and $ -0.3 < u_{1,2} < 0.3$ m s$^{-1}$. Different density ratios are selected, namely $r = 0.98, 0.9, 0.7, 0.5,$ and $0.3$, with $g$ set to $9.8$ m s$^{-2}$.


Note that Roe schemes compute $\bb{Q}$, whereas IFCP and PVM-2U schemes compute only $\bb{C}$, which is additionally multiplied by $\bb{A}$ for comparison. The errors are estimated by a relative bias error given by Eq.~(\ref{eq:BE}), with N-Roe being the reference value.

\begin{figure}[htbp]
	\center
	\includegraphics[width=6cm]{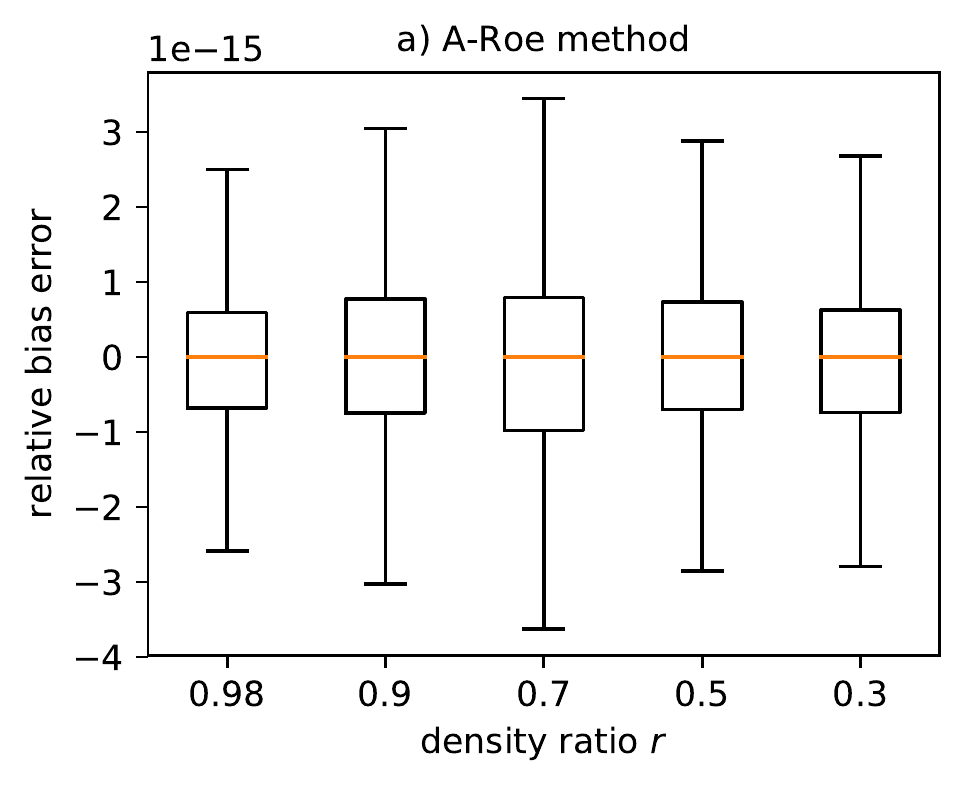}
	\hfill	
	\includegraphics[width=6cm]{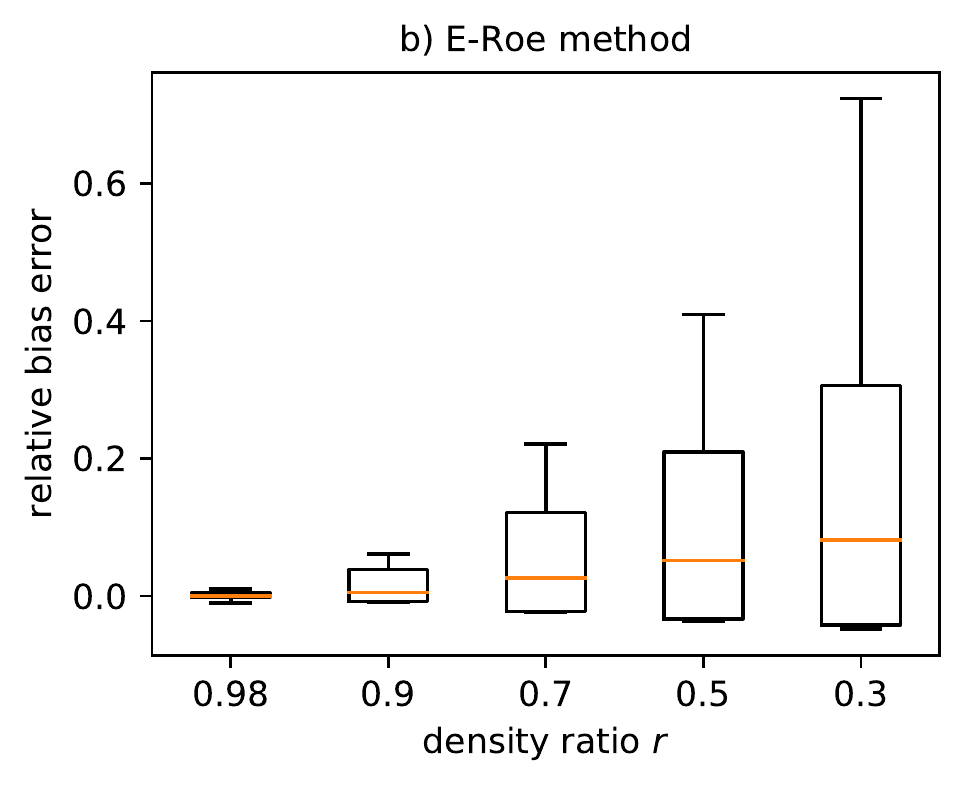}
	\vfill
	\includegraphics[width=6cm]{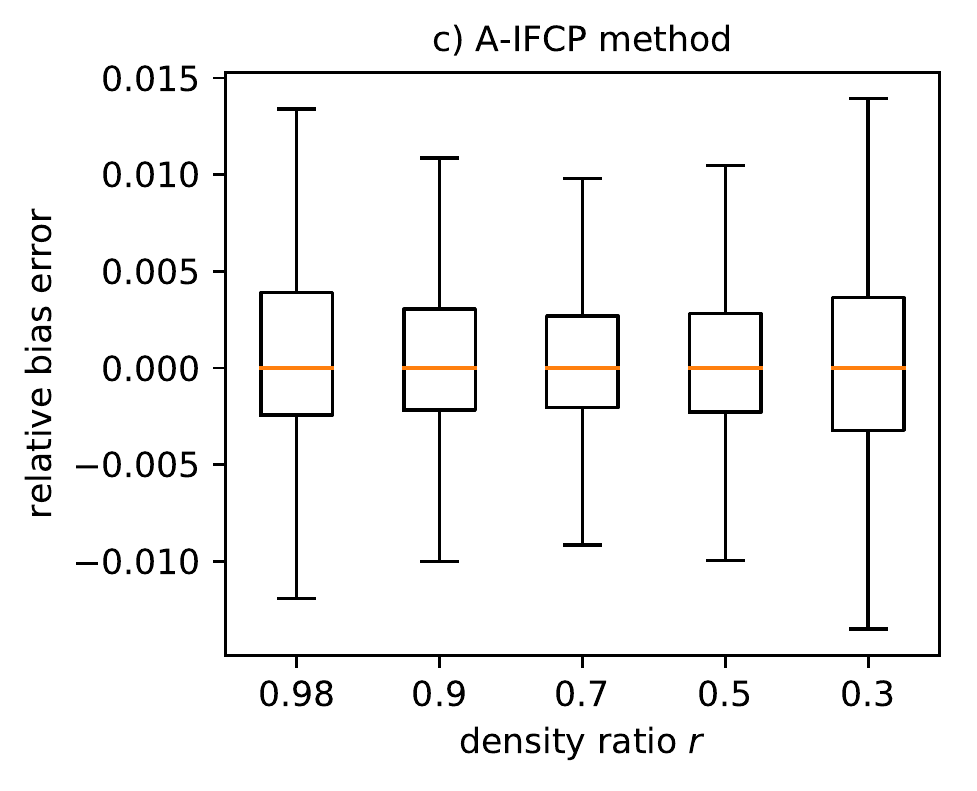}
	\hfill	
	\includegraphics[width=6cm]{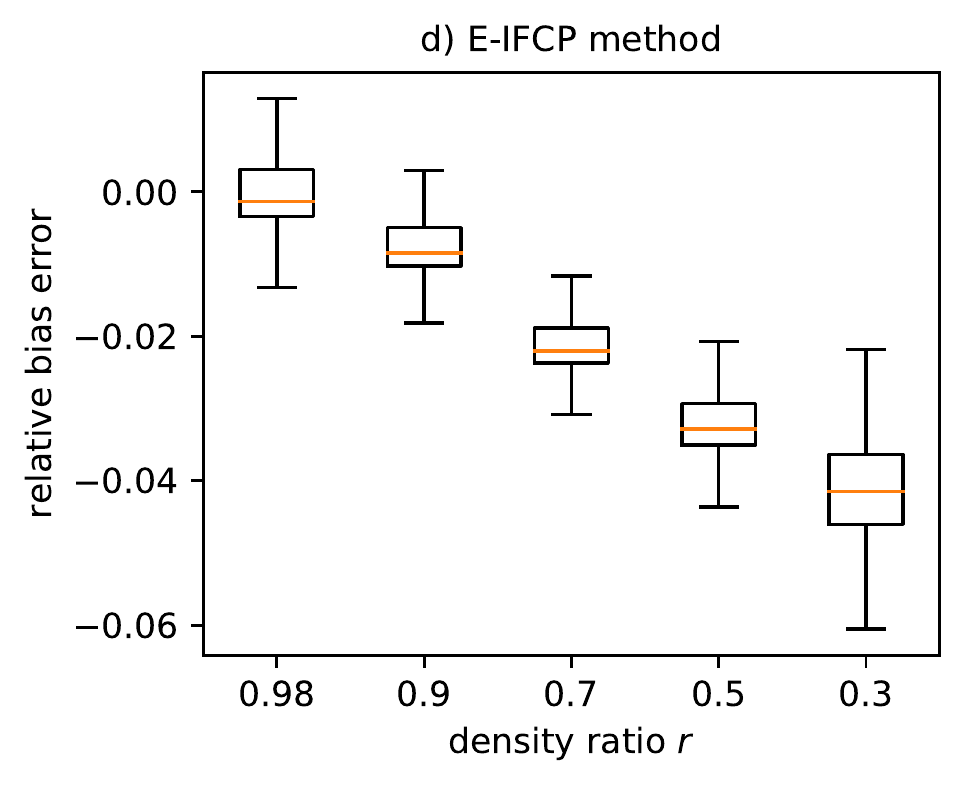}
	\vfill
	\includegraphics[width=6cm]{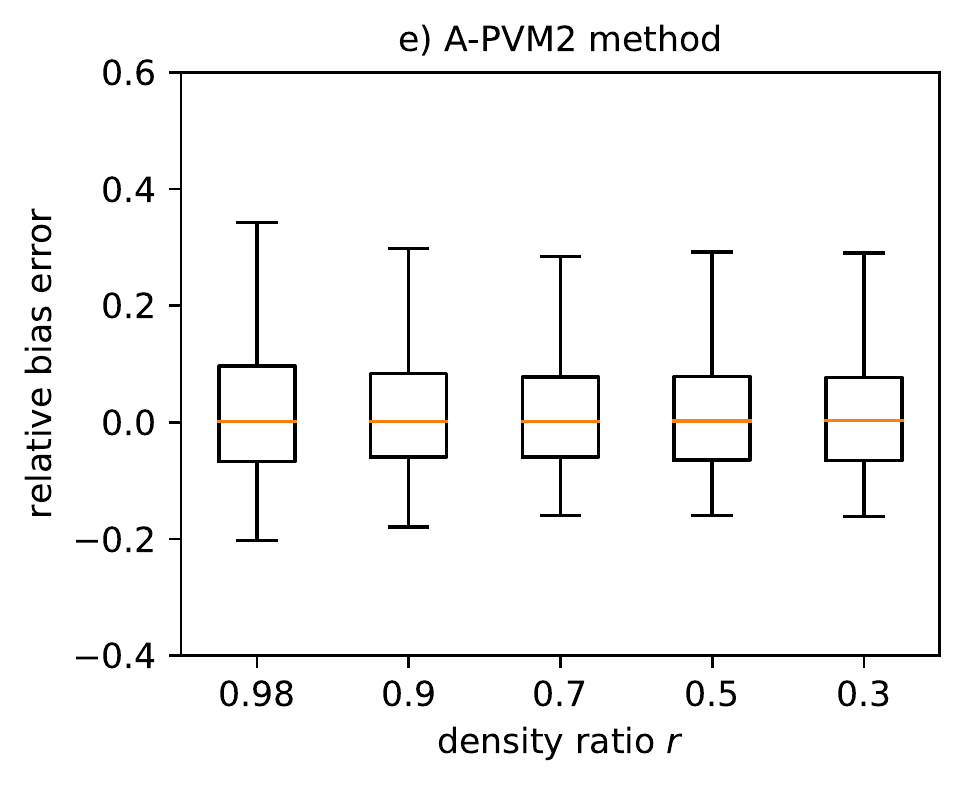}
	\hfill	
	\includegraphics[width=6cm]{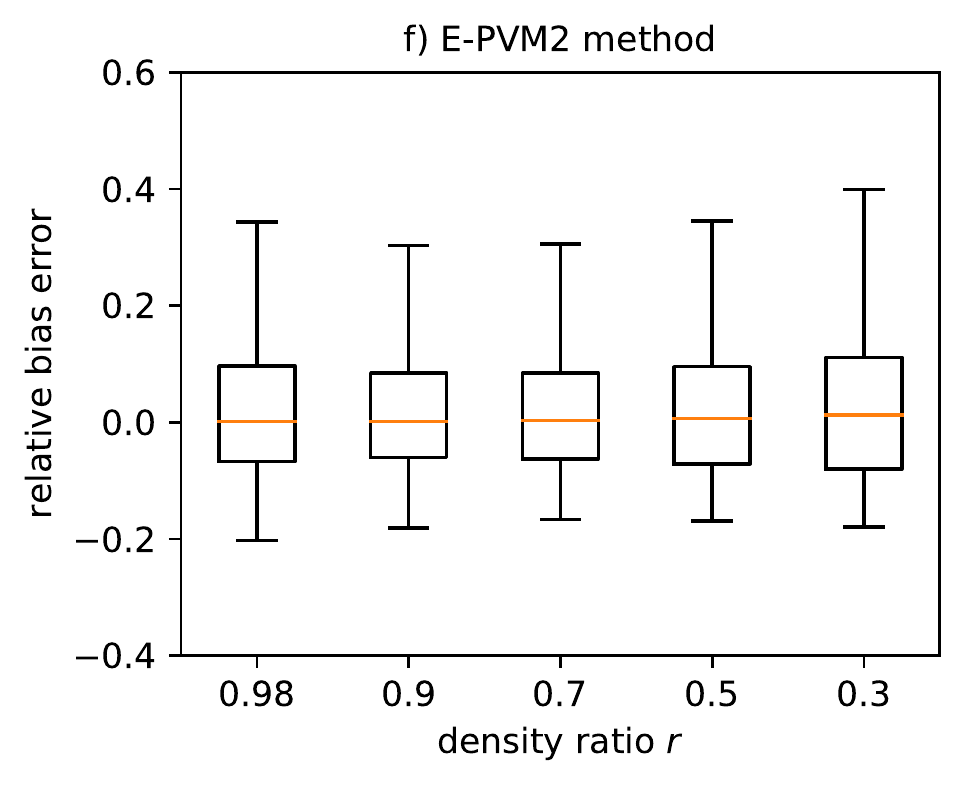}
	\caption{Boxplot of relative bias errors in computing one million numerical viscosity matrices by: a) A-Roe, b) E-Roe, c) A-IFCP, d) E-IFCP, e) A-PVM2, f) E-PVM2. Boxes denote the interquartile range and median value, while whiskers denote min and max values.}
	\label{fig:f_errors}
\end{figure}

Fig.~\ref{fig:f_errors} illustrates the statistical representation of relative bias errors for one million sets of independent flow parameters. Results obtained by A-Roe (Fig.~\ref{fig:f_errors}a) and E-Roe (Fig.~\ref{fig:f_errors}b) schemes correspond to the accuracy of the eigenvalue solvers they use. The average error of the A-Roe method is of the order $10^{-15}$ and the maximum errors are always below $10^{-14}$. On the other hand, the average errors of the E-Roe scheme are much higher and increase with density difference. They amount to $10^{-2}$ for $r=0.98$ and just under $1.0$ for $r=0.3$. Note that both positive and negative errors are observed, although in the case of E-Roe the positive errors (overestimation of the viscosity matrix) are more dominant.

The errors of IFCP and PVM-2U schemes are much higher than the A-Roe method, and comparable to E-Roe scheme. Between the two, PVM-2U (Fig.~\ref{fig:f_errors}e,f) is one order of magnitude less accurate than the IFCP scheme (Fig.~\ref{fig:f_errors}c,d). This emphasizes the importance of using both external and internal eigenvalue information. For A-IFCP schemes relative error is always under $10^{-2}$, whereas for E-IFCP the relative error grows with decreasing $r$. It is interesting to note that E-IFCP, in contrast to E-Roe, underestimates the viscosity matrix. In PVM-2U scheme, the differences between the analytical and approximated eigenvalues are negligible, although the errors generated by approximated eigenvalues are somewhat higher for smaller values of $r$. 

It is interesting to note that the errors for $r<0.7$ are lower in IFCP and PVM-2U schemes (Fig.~\ref{fig:f_errors}d,f) than in the E-Roe scheme (Fig.~\ref{fig:f_errors}b). This is surprising, considering that they both use the same approximated eigenvalue solver, but the Roe scheme uses all eigenvalues, in contrast to IFCP which uses external and one internal eigenavalues, and PVM-2U which only uses minimum and maximum eigenvalues. It seems that the error produced by the E-Eig is somehow compensated by an approximation of the viscosity matrix in PVM-2U schemes. Considering the errors they produce, Roe schemes are very sensitive to the choice of eigenvalues, IFCP schemes are to some degree affected by the eigenvalues, whereas PVM-2U schemes are nearly unaffected by this choice.

The computational time of the schemes are shown in Tab.~\ref{tab:f_erros}. As expected, the slowest scheme is N-Roe due to numerical eigen-decomposition. E-PVM2 and E-IFCP schemes are equally the fastest, followed by A-IFCP and A-PVM2. E-Roe and A-Roe are about 4-5 times faster than N-Roe, and about 35\% slower than corresponding IFCP and PVM-2U schemes. 
When schemes based on A-Eig are computed, they are 0.14 - 0.19 s slower in comparison to the same schemes with E-Eig, which corresponds to the overhead due to analytical eigenstructure solver. Note that the computation of eigenvectors takes as much or even more time than the computation of eigenvalues. This is important because PVM-2U schemes do not require computation of eigenvectors, only eigenvalues. 

The overall overhead from computing the analytical eigenvalues in comparison to approximated ones is only 8-13\%, depending on the scheme. For N-Roe scheme, the eigendecomposition comprises over 80\% of the total CPU time, whereas for A-Roe and E-Roe schemes, the computation of eigenvalues and eigenvectors comprises only 20-26\% of the total CPU time. On the other hand, for IFCP and PVM-2U schemes, computation of approximated eigenvalues comprises about 8\% of CPU time, and computational of closed-form eigenvalues about 18\%. Therefore, the majority of time is used for computing viscosity matrices and not eigenvalues.

\begin{table}[htbp]
	\small
	\centering
	\caption{Total CPU time (including time required only for eigenvalues $\bs{\Lambda}$ and eigenvectors $\bb{K}$) when computing one million numerical viscosity matrices by N-Roe, A-Roe, E-Roe, A-IFCP, E-IFCP, A-PVM2, and E-PVM2. (best of 5 runs)}
	\begin{tabular}{clllllll}
		\hline
		Scheme  & N-Roe & A-Roe & E-Roe & A-IFCP & E-IFCP & A-PVM2 & E-PVM2 \\
		\hline
		$\bs{\Lambda}$ (s)   & 4.13  & 0.29  & 0.11  & 0.29  & 0.11   & 0.29  & 0.11 \\
		$\bb{K}$ (s)   & 3.38  & 0.28  & 0.27  & -  & -   & -  & - \\
		Total (s)   & 9.50  & 2.11  & 1.93  & 1.58  & 1.41   & 1.57  & 1.38 \\
		\hline
	\end{tabular}%
	\label{tab:f_erros}%
\end{table}%

\subsection{Numerical test I: Steady flow over smooth non-flat bed with zero flow rate}

Since all considered schemes (N-Roe, A-Roe, E-Roe, A-IFCP, E-IFCP, A-PVM2, and E-PVM2) are theoretically stable only for exact eigenvalues \cite{castro2001q,fernandez2011intermediate}, the well-balance properties of these schemes with differently computed eigenvalues should be verified in practice.
The first numerical test is designed to test the well-balanced properties of the schemes when simulating water at rest in a channel with smooth non-flat bed. To do so, the spatial domain is set to [0, 1], and a bed function is defined by a single bump:
\begin{equation}
b(x) = 
\begin{cases}
\dfrac{\cos \left( \pi (x-0.5) / 0.1 \right) + 1}{4} \textrm{ m}, \quad \textrm{if } 0.4 < x < 0.6 \textrm{ m} \\
0.0 \textrm{ m}, \quad \textrm{otherwise}
\end{cases}
\end{equation}
The initial condition is given by:
\begin{equation}
h_1(x,0) = 
0.4 \textrm{ m}, 
\quad
h_2(x,0) = 
0.6 \textrm{ m} - b(x)
\end{equation}
\begin{equation}
u_1(x,0) = u_2(x,0) = 0 \textrm{ m s}^{-1}
\end{equation}

Non-reflective conditions are imposed at the boundaries, and the relative density ratio is set to $r=0.98$. 
A single grid density of $\Delta x$ = 1/100 m and a fixed time step $\Delta t=0.002$ s is chosen. 

\begin{figure}[htbp]
	\center
	\includegraphics[width=6cm]{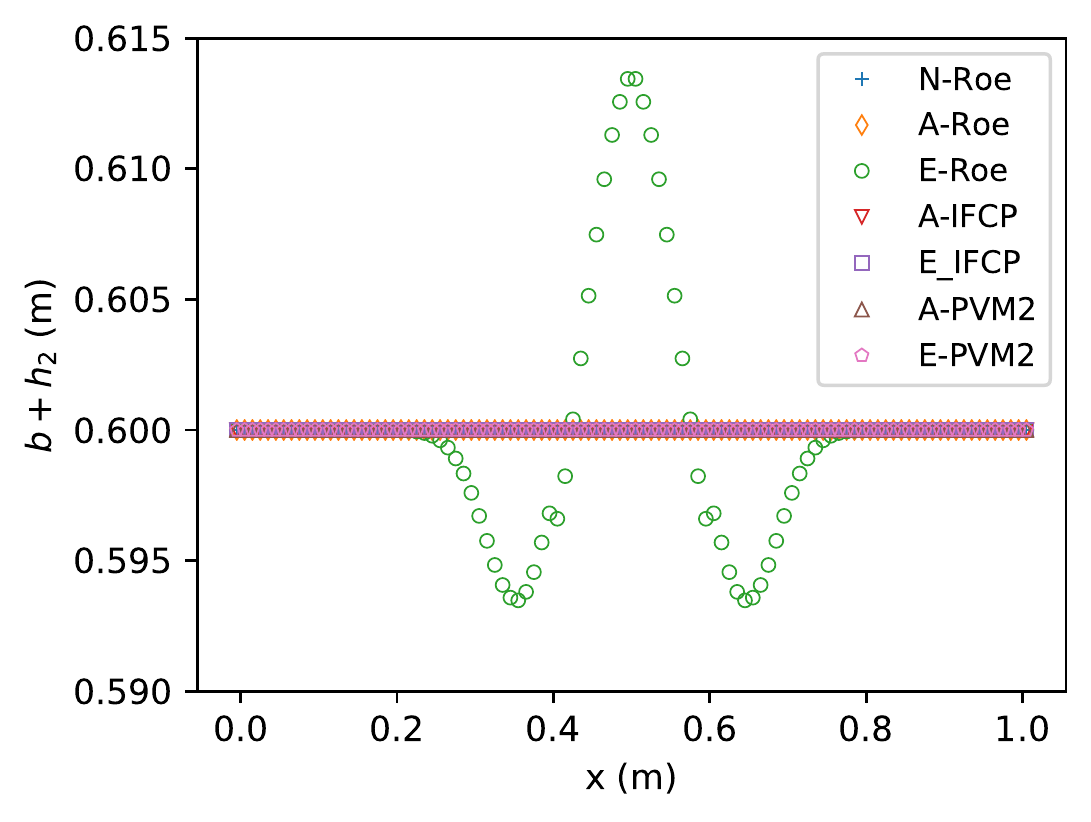}
	\hfill
	\includegraphics[width=6cm]{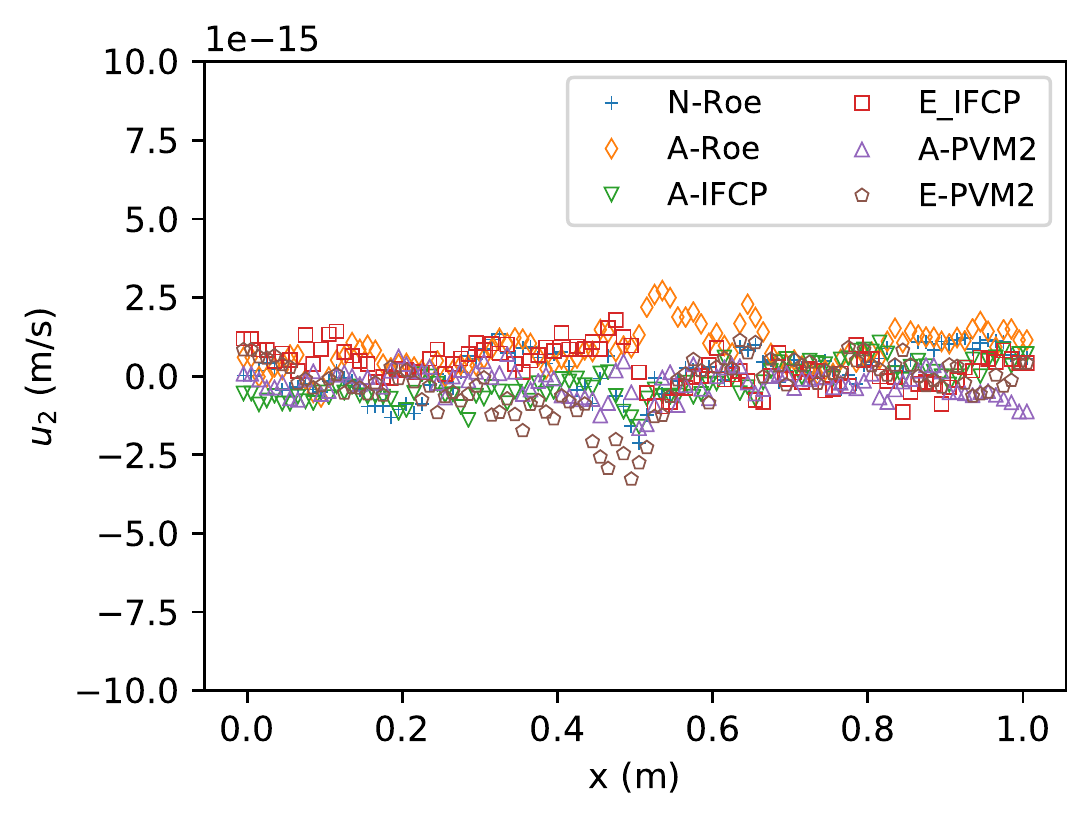}
	\caption{Test I: Computed interface and lower layer velocity obtained by N-Roe, A-Roe, E-Roe, A-IFCP, E-IFCP, A-PVM2, and E-PVM2 scheme at $t=0.1$ s with $\Delta x = 1/100$ m}
	\label{fig:Results_Test00}
\end{figure}

Figure \ref{fig:Results_Test00} shows the interface profile and the lower layer velocities obtained by N-Roe, A-Roe, E-Roe, A-IFCP, E-IFCP, A-PVM2, and E-PVM2 numerical schemes at $t=0.1$ s with $\Delta x = 1/100$ m. All schemes satisfy the well-balance properties for water at rest (the velocities are close to the order of computational precision), except E-Roe, where noticeable spurious oscillations appear at $0.4<x<0.6$ m. Similar results were obtained for other choices of $r$ and $\Delta x$.  Clearly, these oscillations appear because of eigenvalue approximations, and therefore the E-Roe scheme is excluded from further tests.

IFCP and PVM-2U schemes, however, are not prone to spurious oscillations, even when eigenvalues are approximated. Original studies that proposed these methods \cite{fernandez2011intermediate,castro2012class} do not provide any theoretical proof of the stability of IFCP and PVM-2U schemes for approximated eigenvalues, but they do confirm that no stability issues were observed in their extensive tests.

\subsection{Numerical test II: Internal collision of two dam breaks with $r$=0.98}

This test investigates the performance of numerical schemes when simulating a two-layer flow through a rectangular channel with a flat bottom topography. The flow structure is defined by two internal dam-breaks which eventually collide and produce a superimposed wave. A small density difference between the layers is considered, namely $r=0.98$. The spatial domain is set to [0, 100], and the initial condition is given by:
\begin{equation}
h_1(x,0) = 
\begin{cases}
0.8 \textrm{ m}, \quad \textrm{if } 40 < x < 60 \textrm{ m} \\
0.2 \textrm{ m}, \quad \textrm{otherwise}
\end{cases}
\quad
h_2(x,0) = 1.0 m - h_1(x, 0)
\end{equation}
\begin{equation}
u_1(x,0) = u_2(x,0) = 0 \textrm{ m s}^{-1}
\end{equation}

Non-reflective conditions are imposed at the boundaries. 
Several mesh sizes are considered, namely $\Delta x$ = 1, 1/2, 1/4, and 1/8 m. A variable time step $\Delta t$ is evaluated at each step to satisfy $CFL = 0.9$. 
The reference solution is computed using the A-Roe scheme and a dense grid with $\Delta x$ = 1/16 m.

\begin{figure}[htbp]
	\center
	\includegraphics[width=12cm]{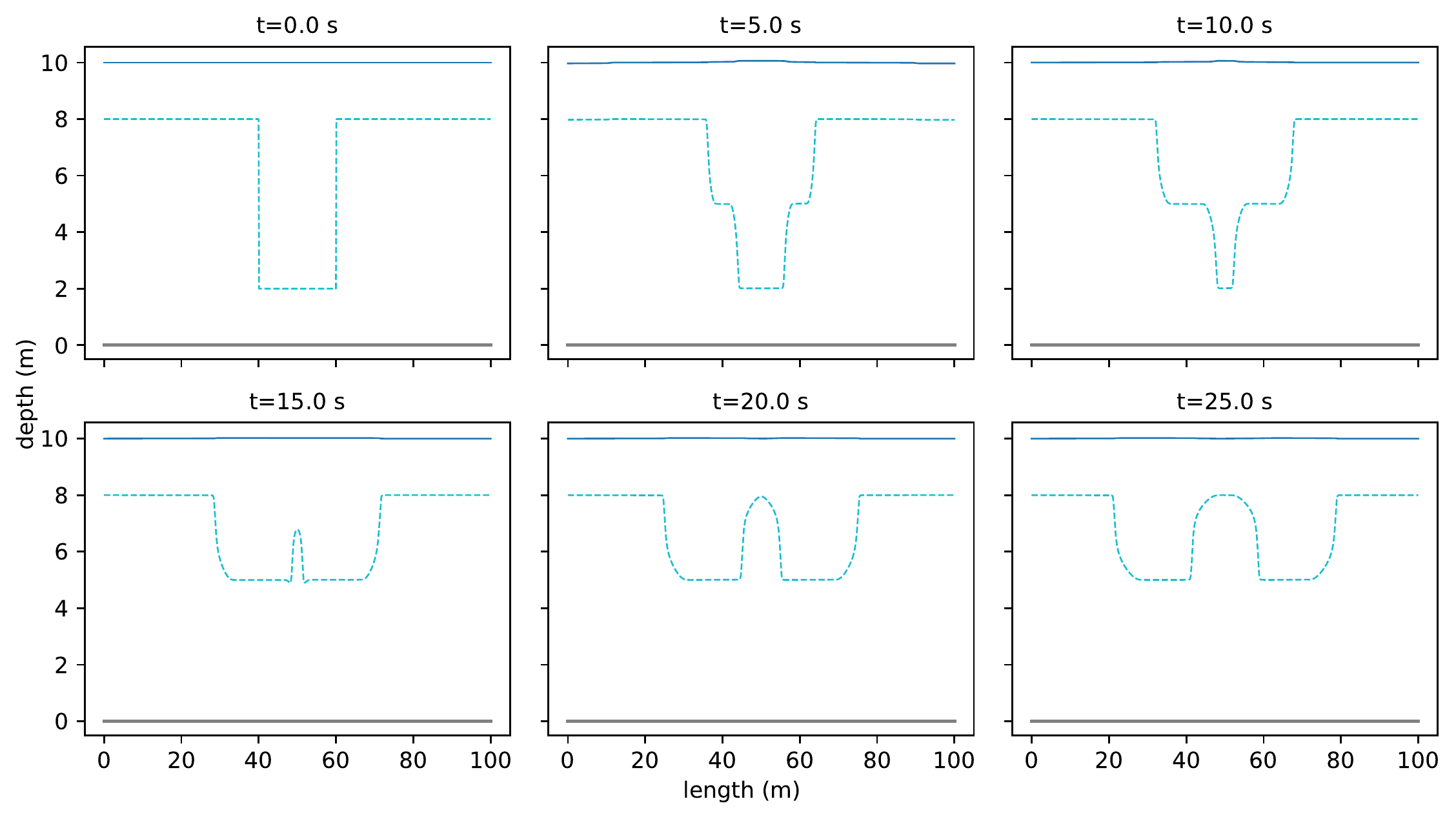}
	\caption{Test II: Temporal evolution of the interface and surface profile (the reference solution)}
	\label{fig:Results_Test3}
\end{figure}

\begin{figure}[htbp]
	\center
	\includegraphics[width=6cm]{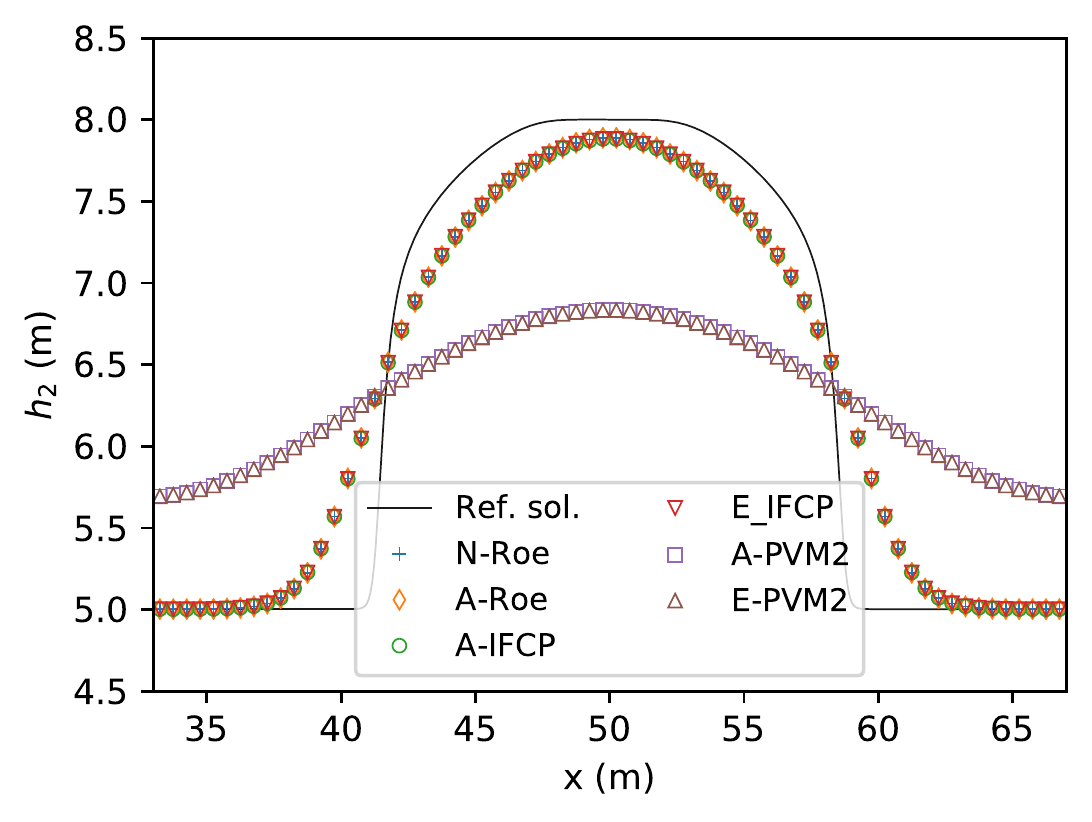}
	\hfill
	\includegraphics[width=6cm]{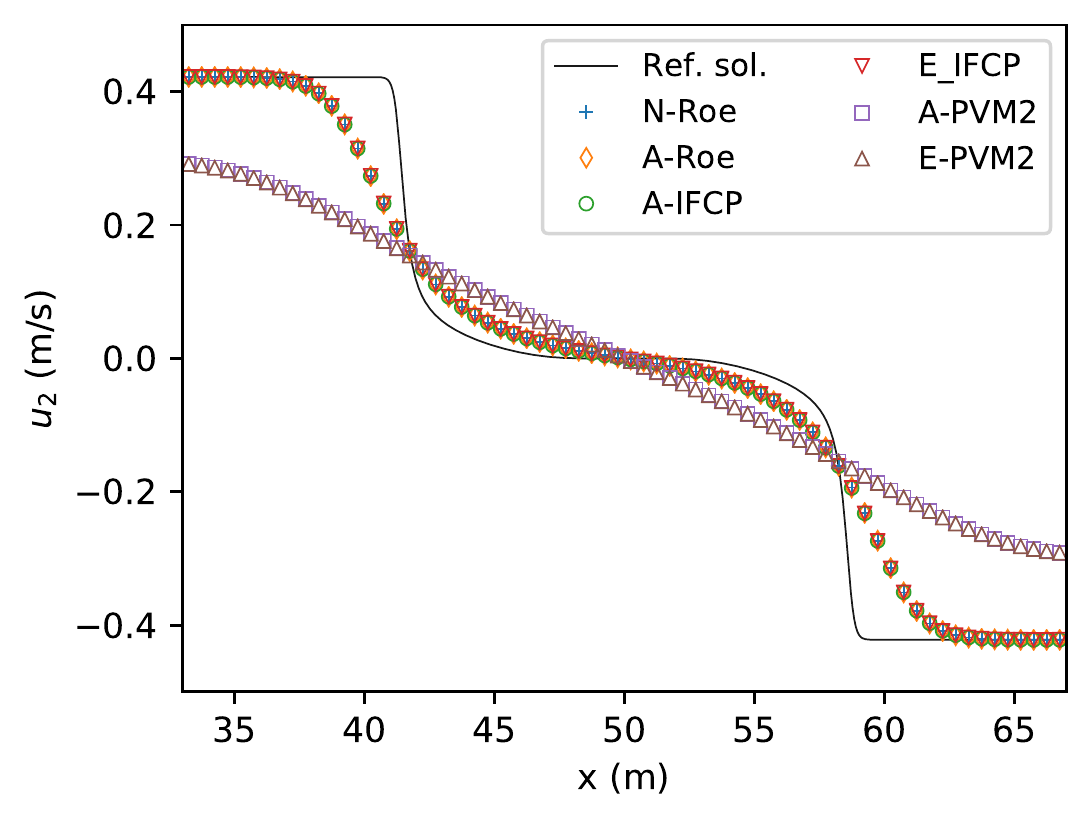}
	\caption{Test II: A detail of the interface depth and lower layer velocity obtained by N-Roe, A-Roe, A-IFCP, E-IFCP, A-PVM2, and E-PVM2 scheme, compared to the reference solution at $t=25$ s with $\Delta x = 0.5$ m}
	\label{fig:Details_Test3}
\end{figure}

Figure \ref{fig:Results_Test3} shows the temporal evolution of the interface and free-surface profiles for the reference solution. Solutions obtained by other methods are not illustrated, because they are almost undistinguishable at this scale. Therefore, a detail of solutions for the interface depth and lower layer velocity are shown in Fig.~\ref{fig:Details_Test3} where N-Roe, A-Roe, A-IFCP, E-IFCP, A-PVM2, and E-PVM2 numerical schemes are compared against the reference solution at $t=25$ s with $\Delta x = 1/2$ m. The results indicate that Roe and IFCP schemes are equally accurate, whereas PVM-2U schemes produce noticeably more diffused results. The implementation of the eigenvalue solver has no apparent influence on the accuracy of the results. This is expected, since approximated eigenvalues exhibit negligible errors for $r=0.98$.

\begin{figure}[htbp]
	\center
	\includegraphics[width=6cm]{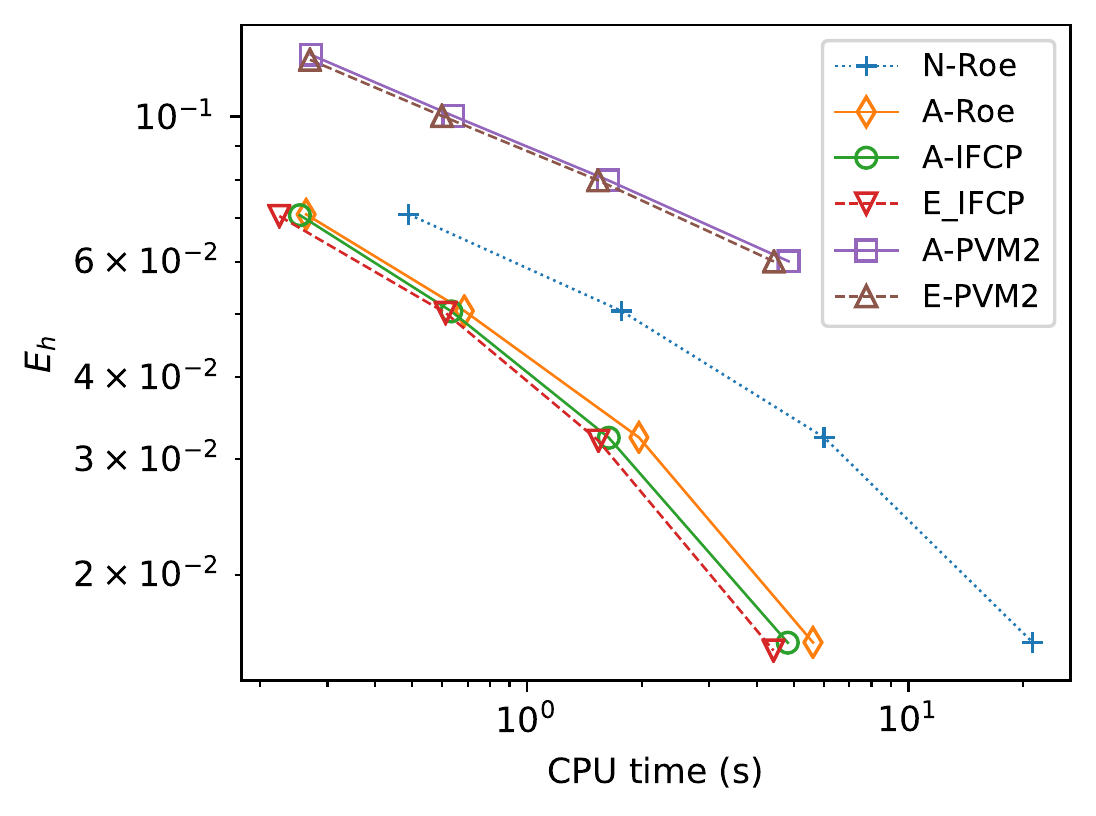}
	\hfill
	\includegraphics[width=6cm]{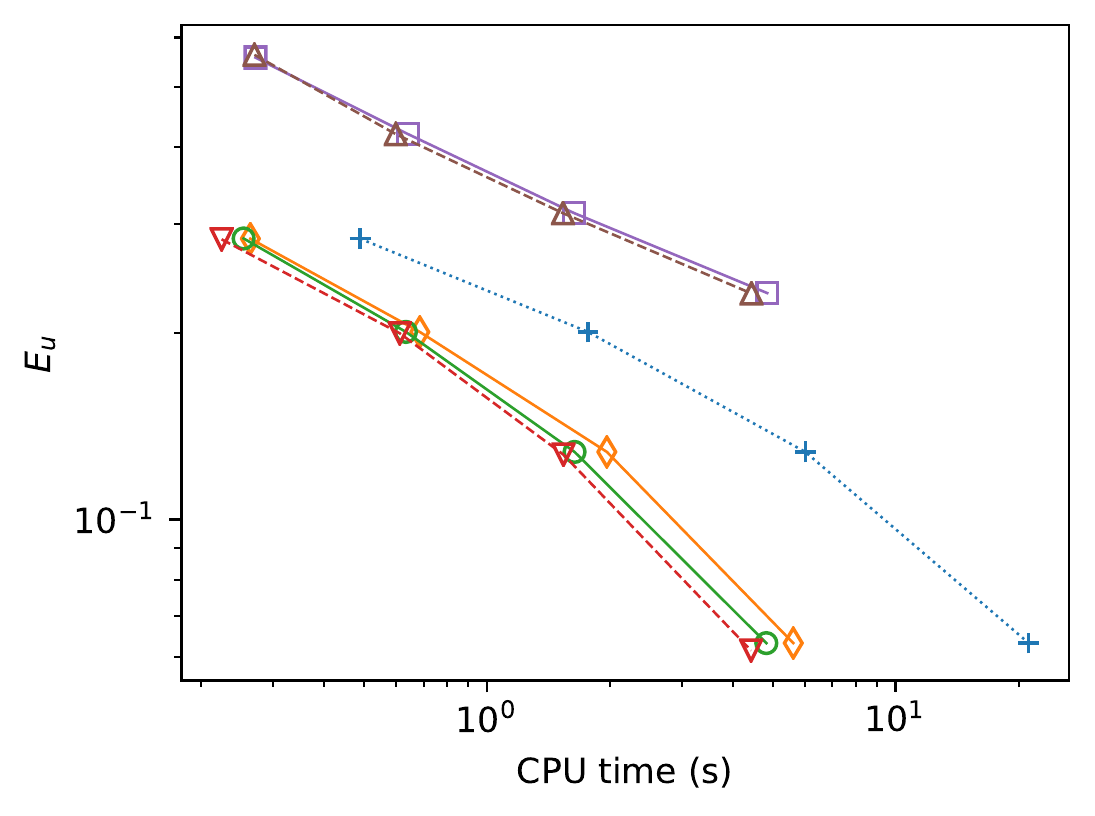}
	\caption{Test II: CPU time vs error for N-Roe, A-Roe, A-IFCP, E-IFCP, A-PVM2, and E-PVM2 scheme, compared to the reference solution}
	\label{fig:Efficiency_Test3}
\end{figure}

A CPU time vs. relative root square error $E_{\Phi}$ is presented in Fig. \ref{fig:Efficiency_Test3}.
The results suggest that the E-IFCP is the most efficient scheme, closely followed by A-IFCP and then A-Roe. In this test, the A-PVM2 and E-PVM2 are the least efficient schemes, even less than N-Roe. For all schemes, the differences in accuracy between the analytical and approximated eigenvalue solvers are insignificant. Approximated eigenvalues produce almost equal results as analytical ones, but need slightly less  computational time. This can be explained by satisfactory accuracy of approximated eigenvalues for $r$ values close to one.

\subsection{Numerical test III: Internal dam break over a sill with $r$=0.98}

This test considers an internal dam-break over uneven bottom, which results in a transition from subcritical to supercritical flow. A small density difference between the layers is considered, namely $r=0.98$. The spatial domain is set to [0, 10], and the bottom topography is defined by a 0.5 m high sill located in the center of the channel, given by the function:
\begin{equation}
b(x) = 0.5 \exp \left( - (x-5)^2 \right).
\end{equation}
The initial condition is given by:
\begin{equation}
h_1(x,0) = 
\begin{cases}
0.2 \textrm{ m}, \quad \textrm{if } x < 5 \textrm{ m} \\
0.8 \textrm{ m}, \quad \textrm{otherwise}
\end{cases}
\quad
h_2(x,0) = 1.5 m - b(x) - h_1(x, 0)
\end{equation}
\begin{equation}
u_1(x,0) = u_2(x,0) = 0 \textrm{ m s}^{-1}
\end{equation}

Non-reflective conditions are imposed at the boundaries. 
Several mesh sizes are considered, namely $\Delta x$ = 1/10, 1/20, 1/40, and 1/80 m. A variable time step $\Delta t$ is evaluated at each step to satisfy $CFL = 0.9$. 
The reference solution is computed using the A-Roe scheme and a dense grid with $\Delta x$ = 1/160 m.

\begin{figure}[htbp]
	\center
	\includegraphics[width=12cm]{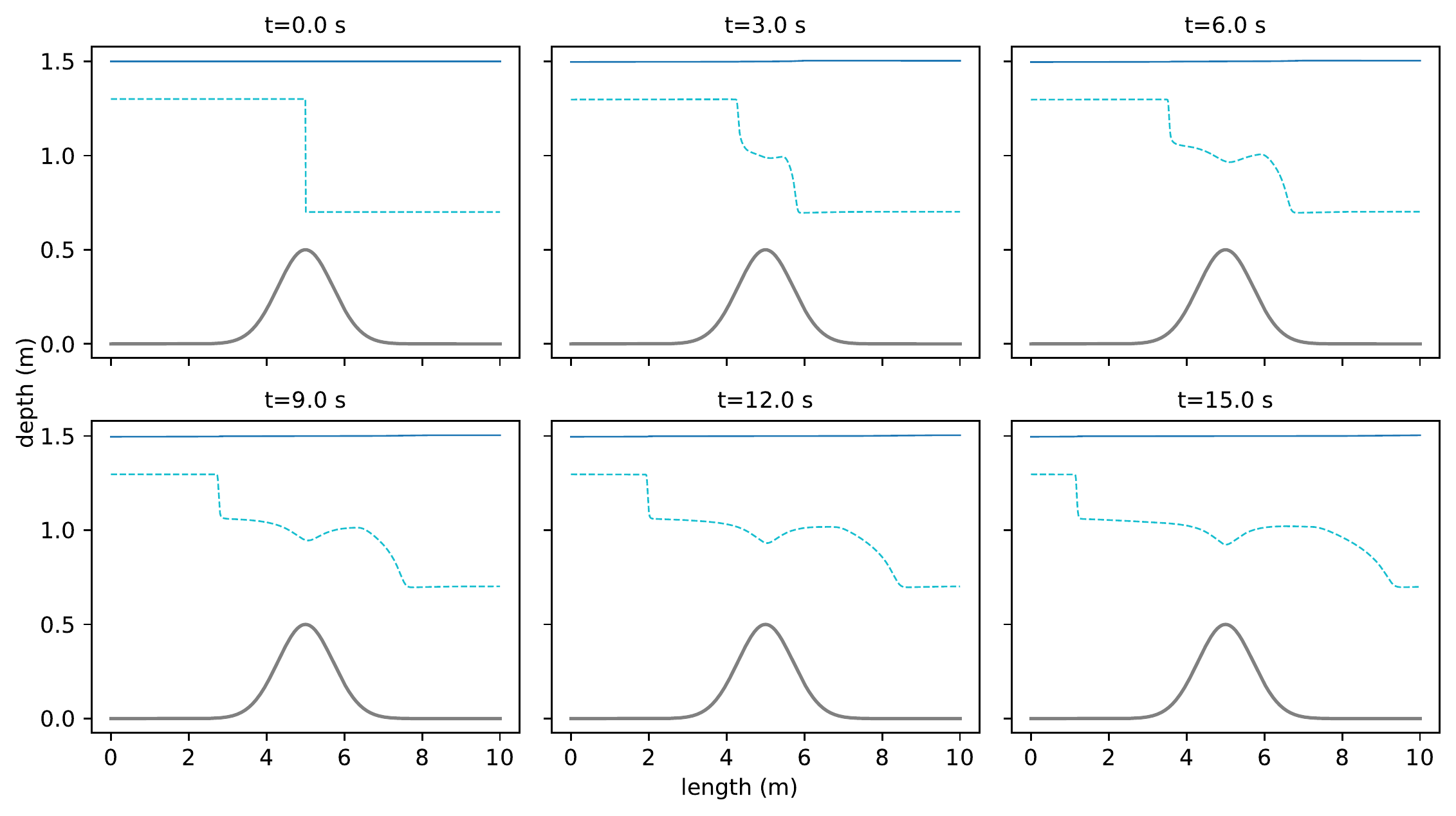}
	\caption{Test III: Temporal evolution of the interface and surface profile (the reference solution)}
	\label{fig:Results_Test4}
\end{figure}

\begin{figure}[htbp]
	\center
	\includegraphics[width=6cm]{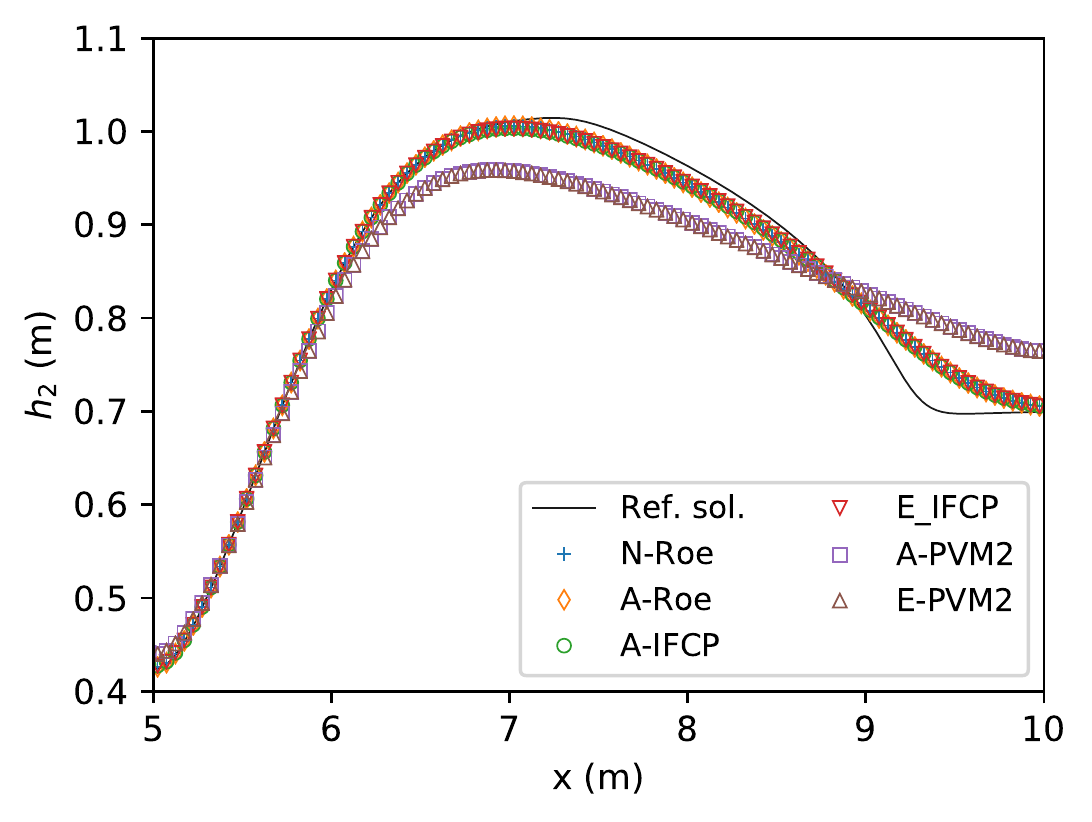}
	\hfill
	\includegraphics[width=6cm]{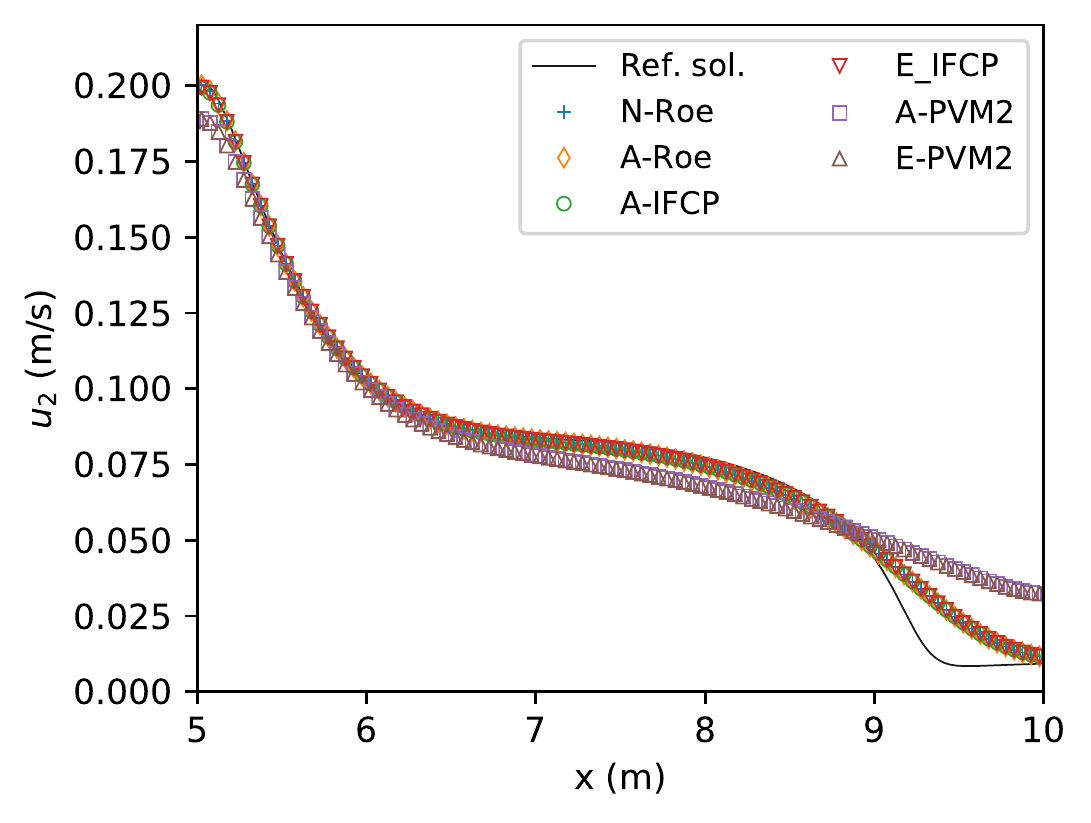}
	\caption{Test III: A detail of the interface depth and lower layer velocity obtained by N-Roe, A-Roe, A-IFCP, E-IFCP, A-PVM2, and E-PVM2 scheme, compared to the reference solution at $t=15$ s with $\Delta x = 0.05$ m}
	\label{fig:Details_Test4}
\end{figure}

Figure \ref{fig:Results_Test4} shows the temporal evolution of the interface and free-surface profiles for the reference solution. A detail of solutions for the interface depth and lower layer velocity are shown in Fig.~\ref{fig:Details_Test4} where N-Roe, A-Roe, A-IFCP, E-IFCP, A-PVM2, and E-PVM2 numerical schemes are compared against the reference solution at $t=15$ s with $\Delta x = 1/20$ m. The results indicate that Roe and IFCP schemes are equally accurate, whereas PVM-2U schemes produce noticeably more diffused results. Same as in the previous case, the implementation of the eigenvalue solver has no apparent influence on the accuracy of the results, since approximated eigenvalues exhibit negligible errors for $r=0.98$.

\begin{figure}[htbp]
	\center
	\includegraphics[width=6cm]{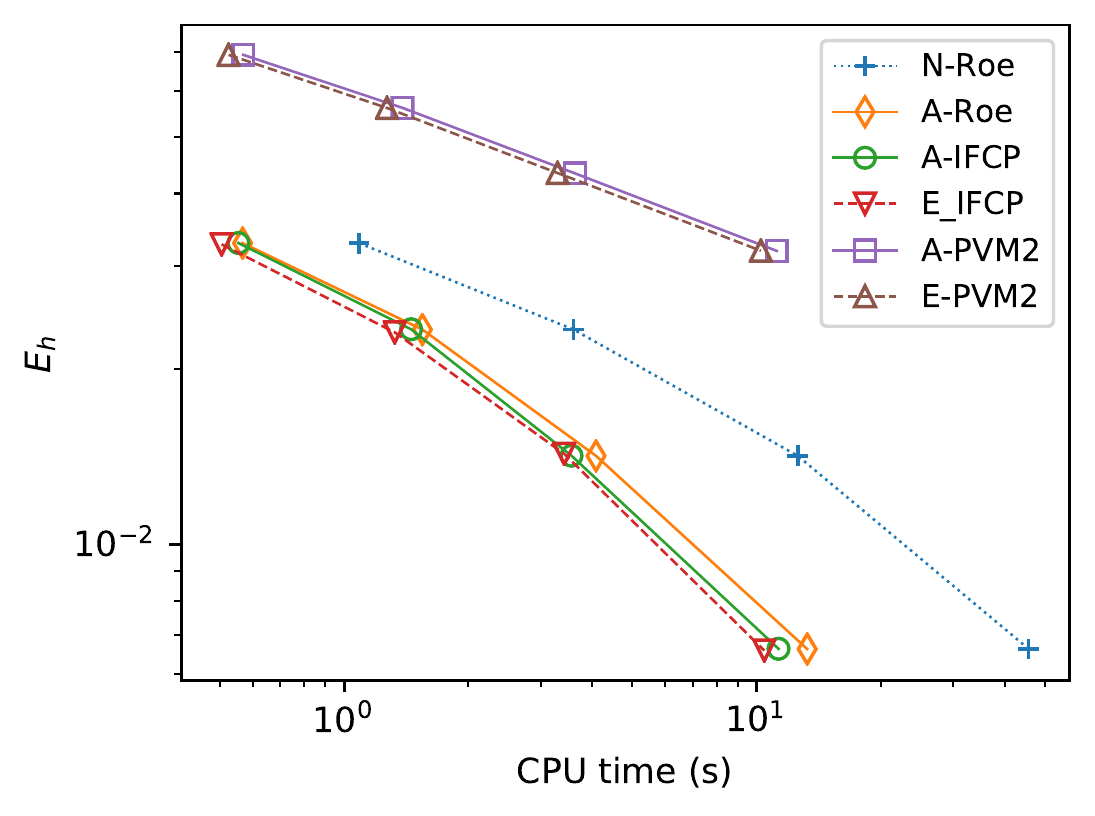}
	\hfill
	\includegraphics[width=6cm]{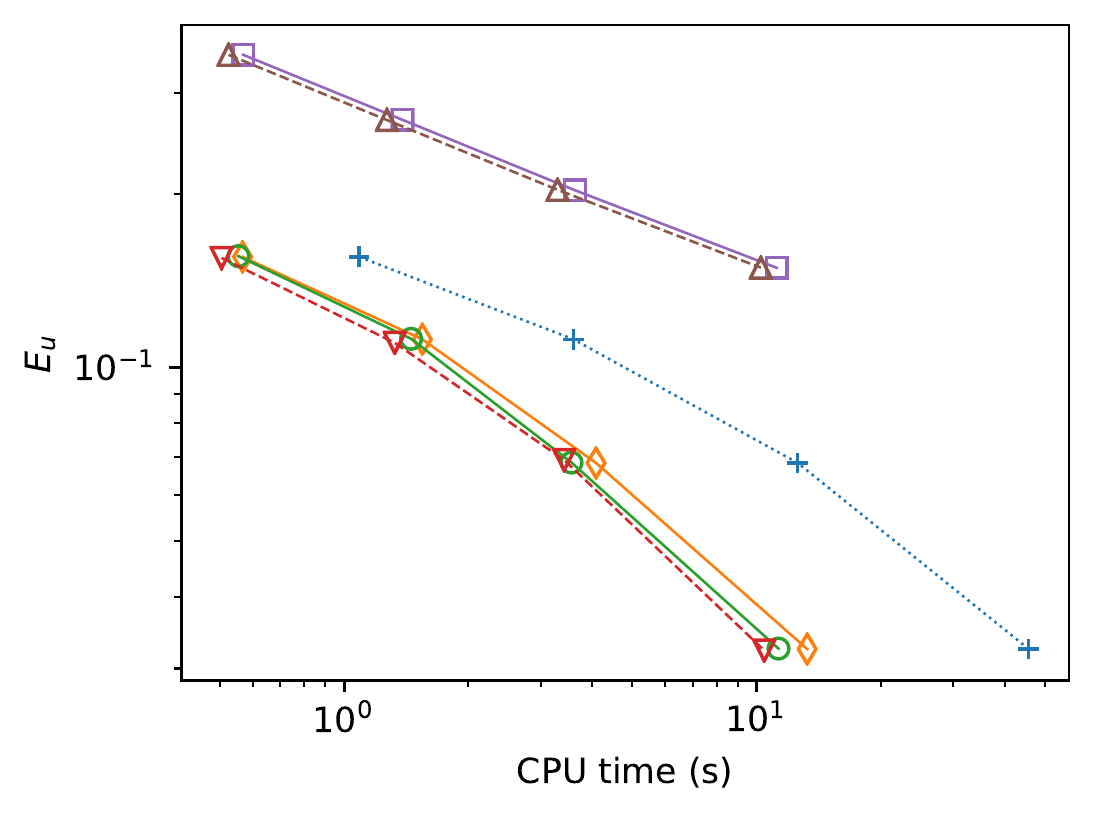}
	\caption{Test III: CPU time vs error for N-Roe, A-Roe, A-IFCP, E-IFCP, A-PVM2, and E-PVM2 scheme, compared to the reference solution}
	\label{fig:Efficiency_Test4}
\end{figure}

A CPU time vs. relative root square error $E_{\Phi}$ is presented in Fig. \ref{fig:Efficiency_Test4}.
The results are very similar to the previous test case, shown in Fig. \ref{fig:Efficiency_Test3}. The E-IFCP is the most efficient scheme, closely followed by A-IFCP and then A-Roe. The A-PVM2 and E-PVM2 are the least efficient schemes. The differences in efficiency between the analytical and approximated eigenvalue solvers are insignificant for IFCP and PVM-2U schemes. 

\subsection{Numerical test IV: Internal dam break with $r$=0.4}

A two-layer flow through a rectangular channel with a flat bottom topography is considered again. In contrast to a similar internal dam-break scenario presented in \cite{krvavica2018analytical}, here we consider a much larger density difference between the layers, namely $r=0.4$. The spatial domain is set to [0, 50], and the initial condition is given by:
\begin{equation}
h_1(x, 0) = 
\begin{cases}
0.2 \textrm{ m}, \quad \textrm{if } x < 25 \textrm{ m} \\
0.8 \textrm{ m}, \quad \textrm{otherwise}
\end{cases}
\quad
h_2(x,0) = 1.0 m - h_1(x, 0)
\end{equation}
\begin{equation}
u_1(x,0) = u_2(x,0) = 0 \textrm{ m s}^{-1}
\end{equation}

Non-reflective conditions are imposed at the boundaries. 
Several mesh sizes are considered, namely $\Delta x$ = 1/2, 1/4, 1/8, and 1/16 m. A variable time step $\Delta t$ is evaluated at each step to satisfy $CFL = 0.9$. 
The reference solution is computed using the A-Roe scheme and a dense grid with $\Delta x$ = 1/32 m.

\begin{figure}[htbp]
	\center
	\includegraphics[width=12cm]{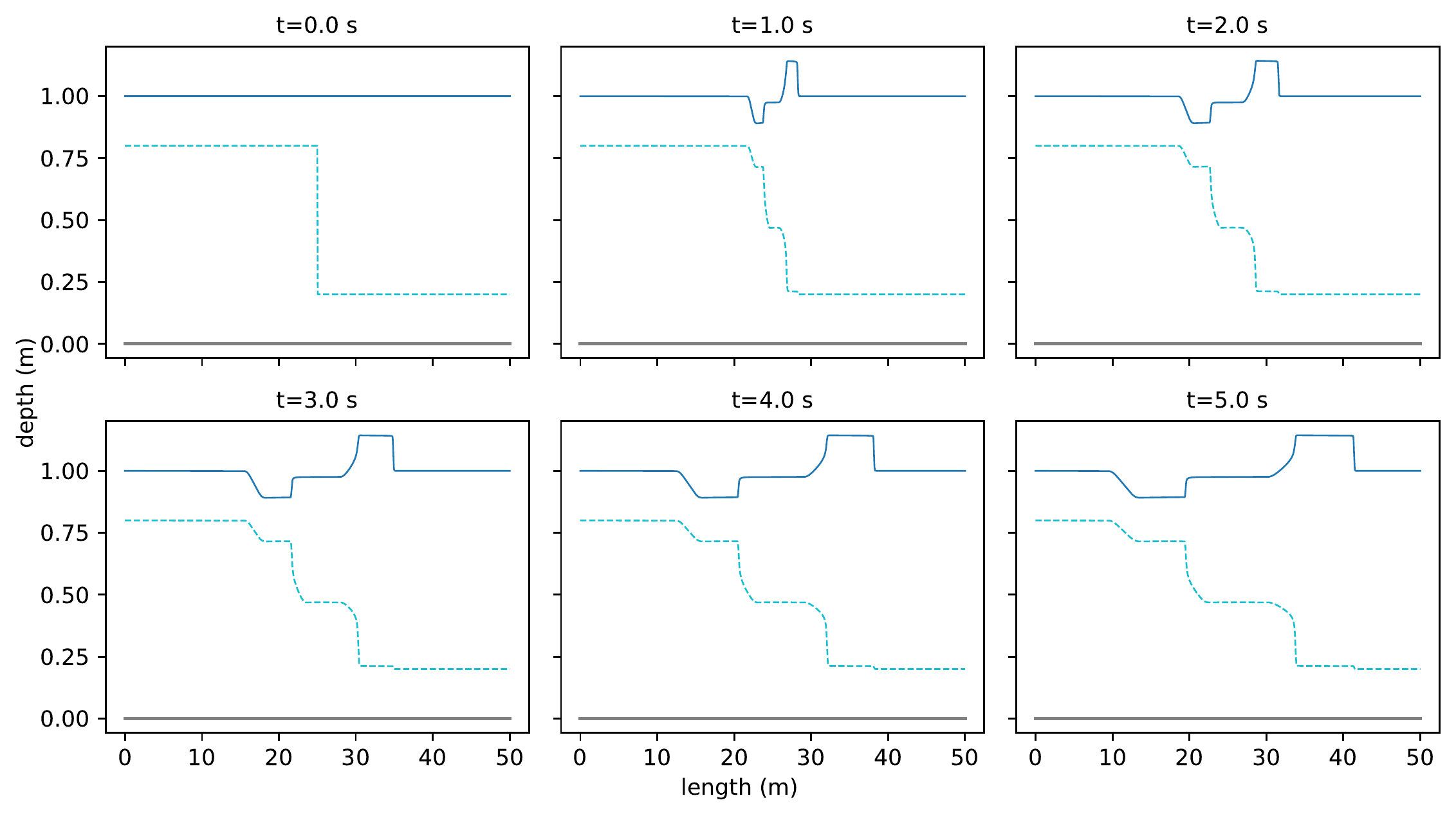}
	\caption{Test IV: Temporal evolution of the interface and surface profile (the reference solution)}
	\label{fig:Results_Test2}
\end{figure}

\begin{figure}[htbp]
	\center
	\includegraphics[width=6cm]{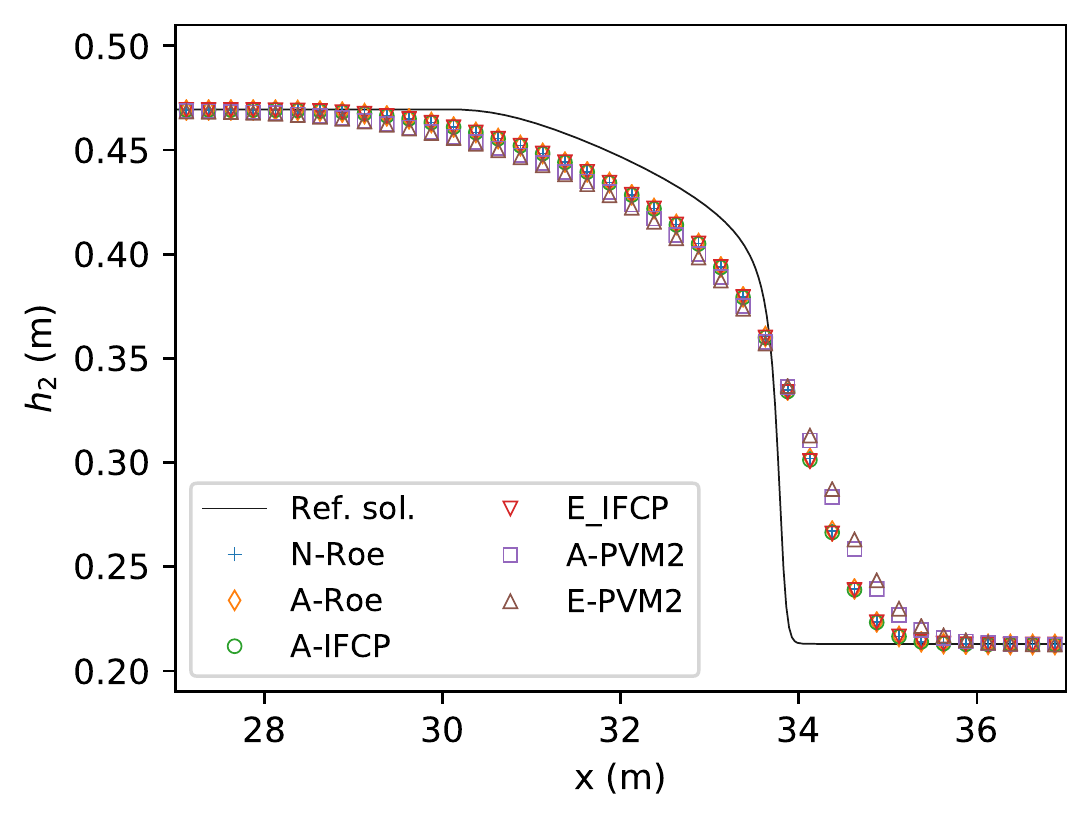}
	\hfill
	\includegraphics[width=6cm]{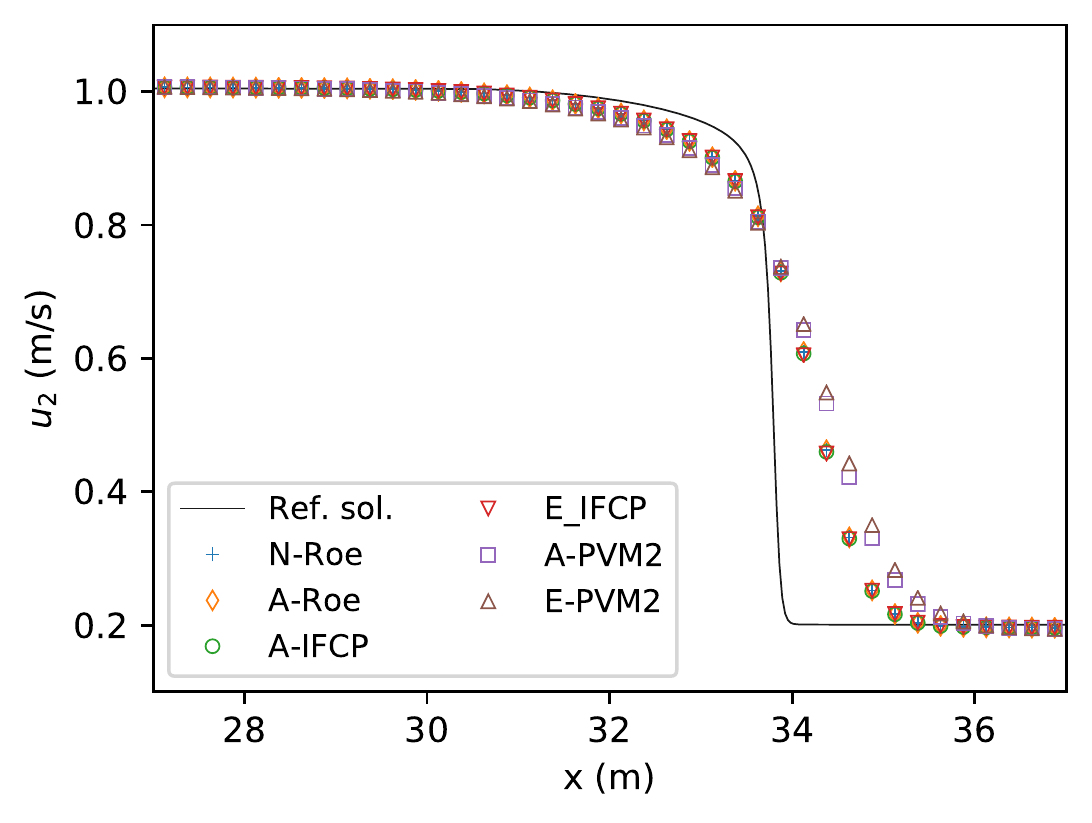}
	\caption{Test IV: A detail of the interface depth and lower layer velocity obtained by N-Roe, A-Roe, A-IFCP, E-IFCP, A-PVM2, and E-PVM2 scheme, compared to the reference solution at $t=5$ s with $\Delta x = 0.25$ m}
	\label{fig:Details_Test2}
\end{figure}

Figure \ref{fig:Results_Test2} shows the temporal evolution of the interface and free-surface profiles for the reference solution. A detail of solutions for the interface depth and lower layer velocity are shown in Fig.~\ref{fig:Details_Test2} where N-Roe, A-Roe, A-IFCP, E-IFCP, A-PVM2, and E-PVM2 numerical schemes are compared against the reference solution at $t=5$ s with $\Delta x = 1/4$ m. The results indicate that all Roe and IFCP schemes are equally accurate, whereas PVM-2U schemes produce more diffused results. Analytical eigenvalues show slightly more accurate values in comparison to the approximated ones, which can be expected since the approximations are derived for $r \approx 1$, which is not the case here.

\begin{figure}[htbp]
	\center
	\includegraphics[width=6cm]{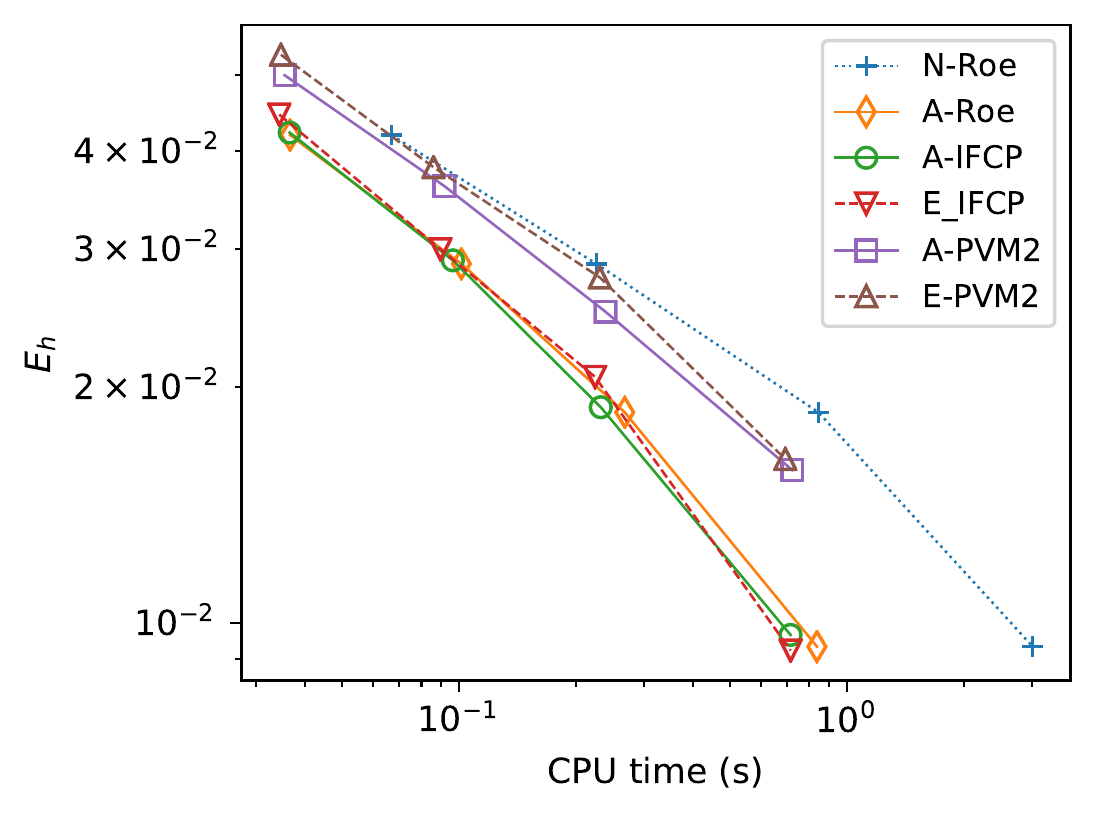}
	\hfill
	\includegraphics[width=6cm]{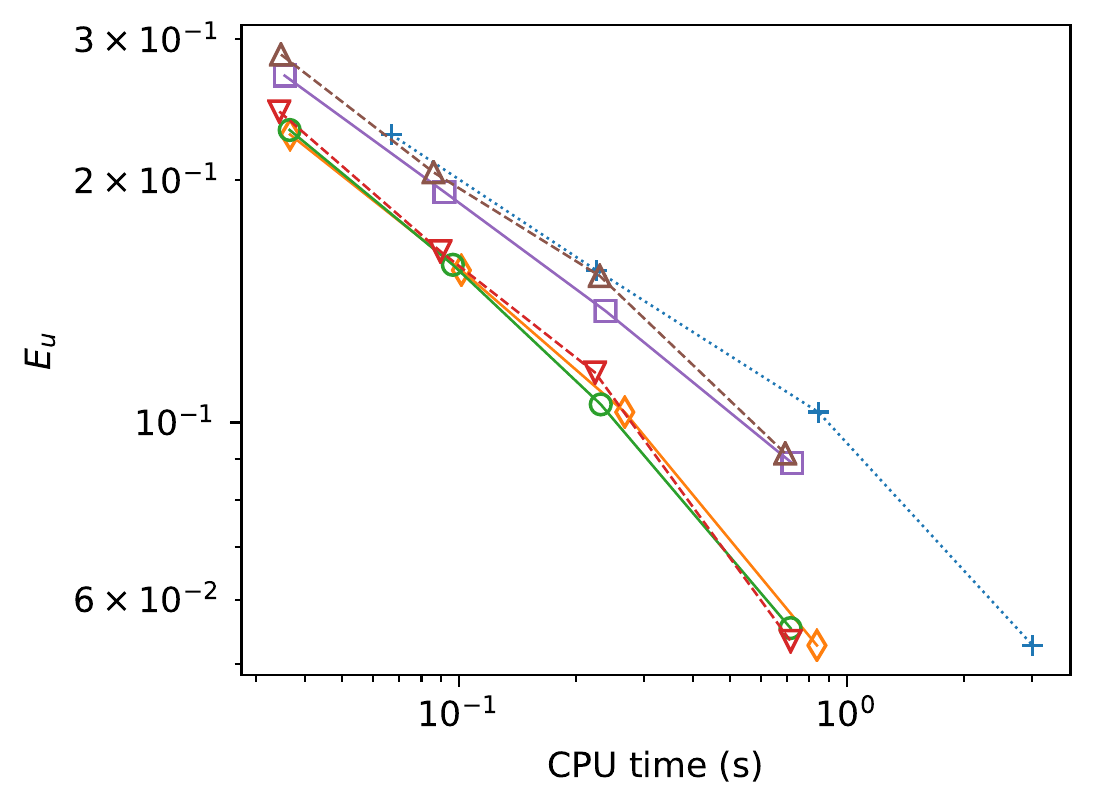}
	\caption{Test IV: CPU time vs error for N-Roe, A-Roe, A-IFCP, E-IFCP, A-PVM2, and E-PVM2 scheme, compared to the reference solution}
	\label{fig:Efficiency_Test2}
\end{figure}

A CPU time vs. relative root square error $E_{\Phi}$ is presented in Fig. \ref{fig:Efficiency_Test2}.
In general, the results show that the A-IFCP, E-IFCP and A-Roe are the most efficient schemes. N-Roe and PVM-2U are noticeably less efficient. The differences in efficiency between the analytical and approximated eigenvalue solvers are not significant, but analytical eigensolvers seem to have a slight advantage over the approximated eigenvalues. Analytical eigenvalues produce more accurate results with a minor increase in the computational cost. 

\subsection{Numerical test V: Internal column collapse on a slope with $r$=0.4}

This test presents a lower-layer column collapse on a sloped bottom. Again a larger density difference between the layers is considered, namely $r=0.4$. The spatial domain is set to [0, 40], and the initial condition is given by:
\begin{equation}
h_1(x, 0) = 
\begin{cases}
0.2 \textrm{ m}, \quad \textrm{if } x < 25 \textrm{ m} \\
0.8 \textrm{ m}, \quad \textrm{otherwise}
\end{cases}
\quad
h_2(x,0) = 1.0 m - h_1(x, 0)
\end{equation}
\begin{equation}
u_1(x,0) = u_2(x,0) = 0 \textrm{ m s}^{-1}
\end{equation}

Non-reflective conditions are imposed at the boundaries. 
Several mesh sizes are considered, namely $\Delta x$ = 2/5, 1/5, 1/10, and 1/20 m. A variable time step $\Delta t$ is evaluated at each step to satisfy $CFL = 0.9$. 
The reference solution is computed using the A-Roe scheme and a dense grid with $\Delta x$ = 1/40 m.

\begin{figure}[htbp]
	\center
	\includegraphics[width=12cm]{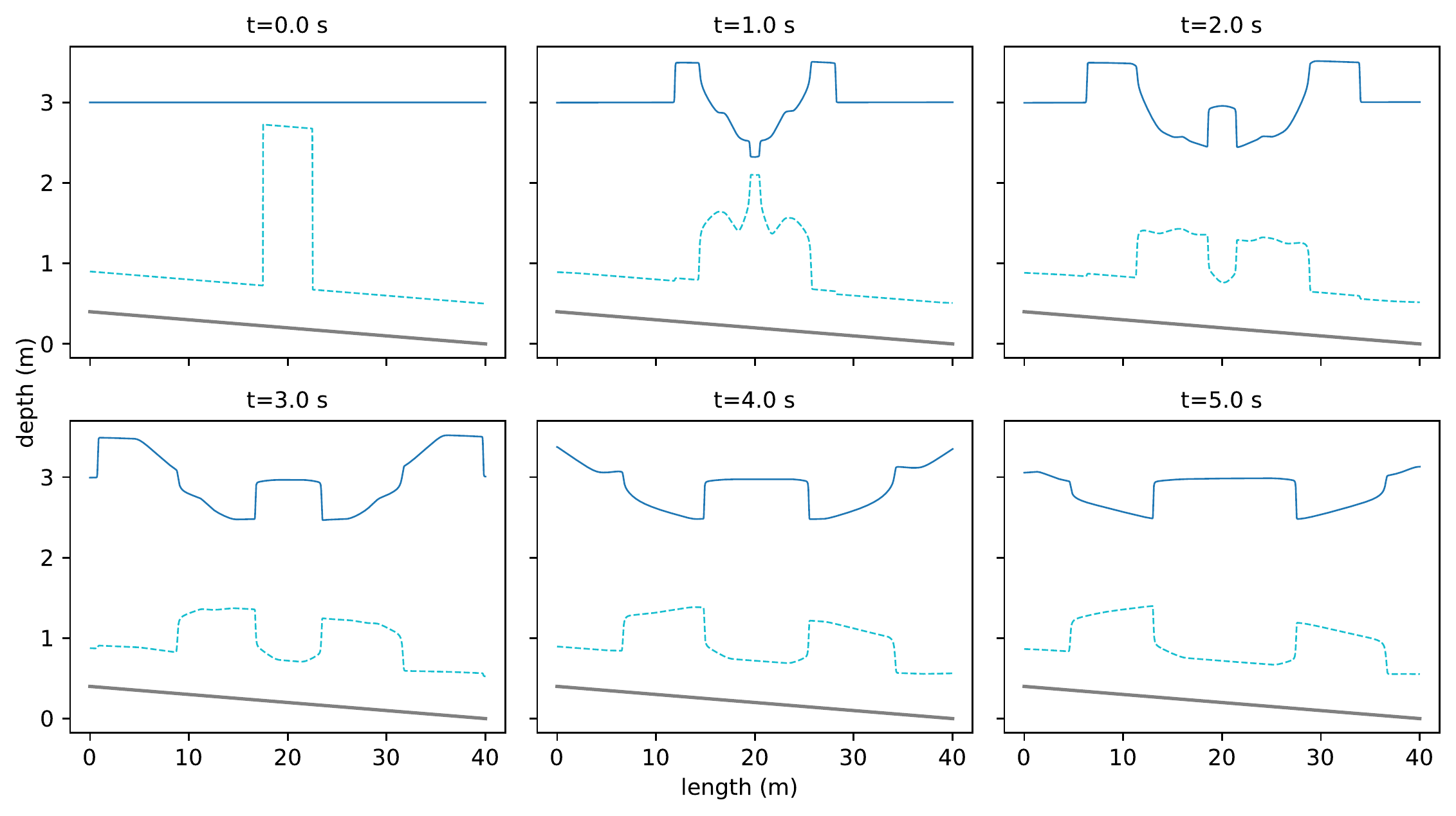}
	\caption{Test V: Temporal evolution of the interface and surface profile (the reference solution)}
	\label{fig:Results_Test5}
\end{figure}

\begin{figure}[htbp]
	\center
	\includegraphics[width=6cm]{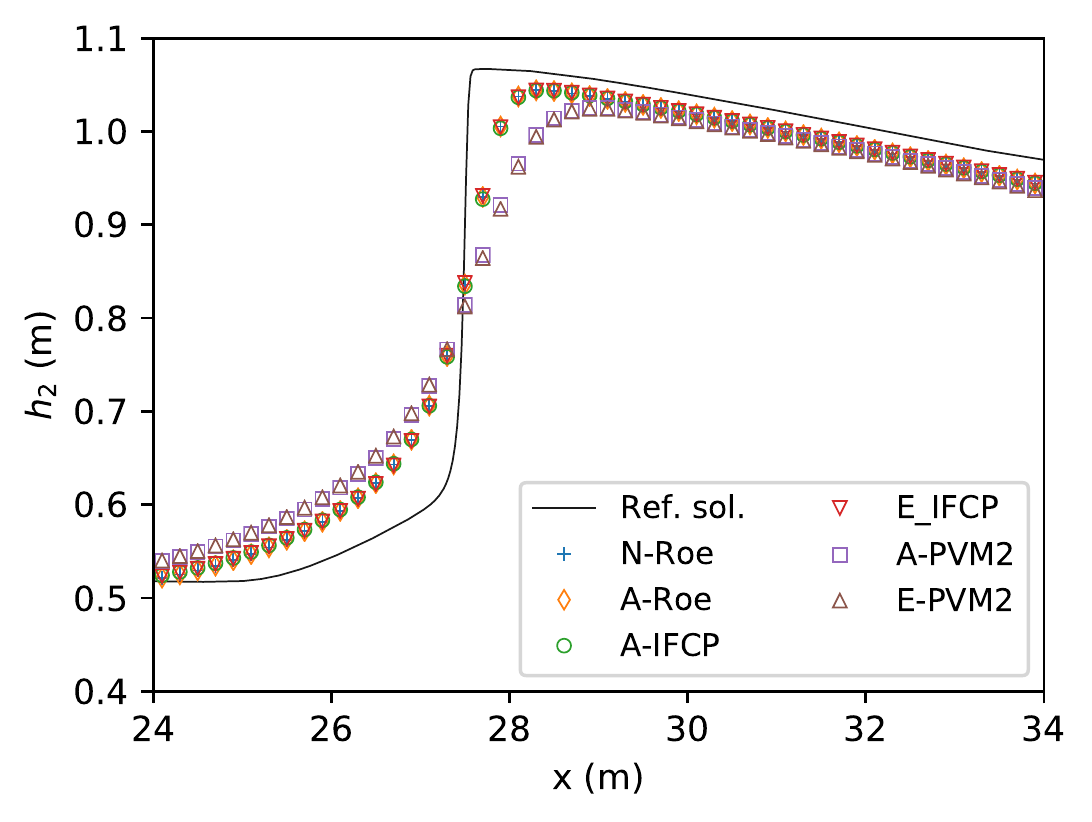}
	\hfill
	\includegraphics[width=6cm]{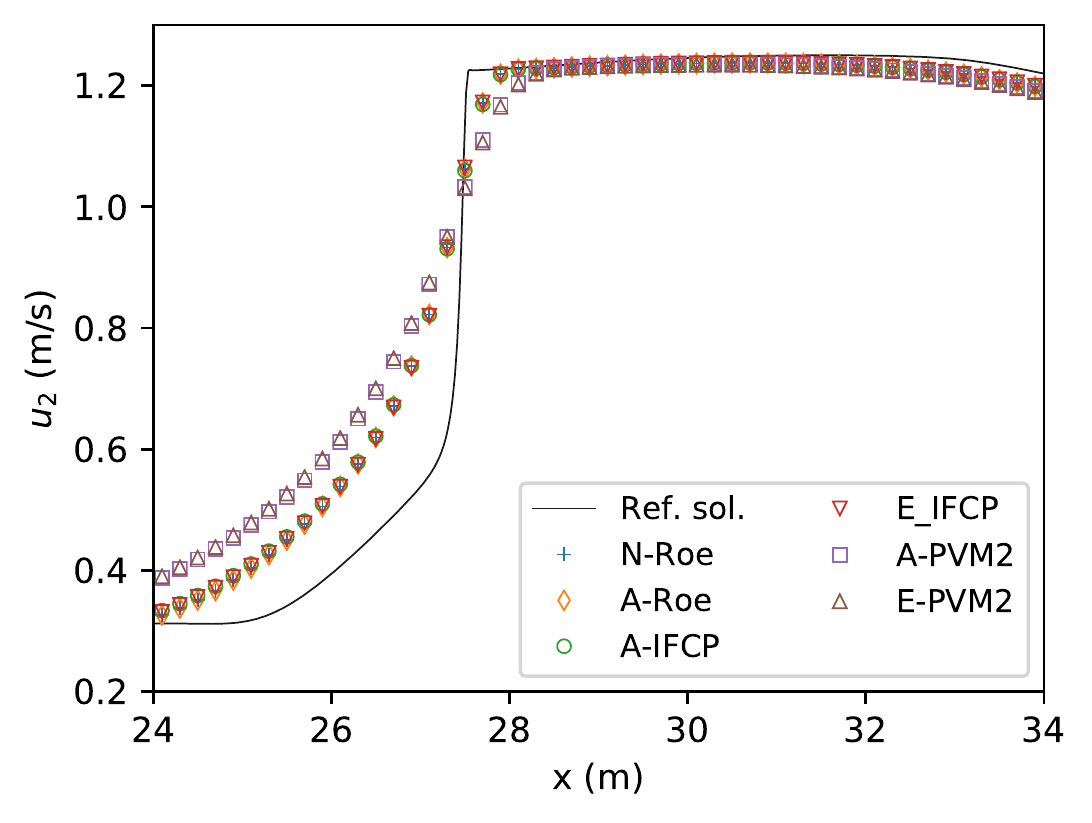}
	\caption{Test V: A detail of the interface depth and lower layer velocity obtained by N-Roe, A-Roe, A-IFCP, E-IFCP, A-PVM2, and E-PVM2 scheme, compared to the reference solution at $t=5$ s with $\Delta x = 0.2$ m}
	\label{fig:Details_Test5}
\end{figure}

Figure \ref{fig:Results_Test5} shows the temporal evolution of the interface and free-surface profiles for the reference solution. A detail of solutions for the interface depth and lower layer velocity are shown in Fig.~\ref{fig:Details_Test5} where N-Roe, A-Roe, A-IFCP, E-IFCP, A-PVM2, and E-PVM2 numerical schemes are compared against the reference solution at $t=5$ s with $\Delta x = 1/5$ m. The results indicate that all Roe and IFCP schemes are equally accurate, whereas PVM-2U schemes produce more diffused results. As in the previous example, analytical eigenvalues show slightly more accurate values in comparison to the approximated ones.

A CPU time vs. relative root square error $E_{\Phi}$ is presented in Fig. \ref{fig:Efficiency_Test2}.
Similar to the previous test, E-IFCP and A-IFCP are the most efficient schemes, closely followed by A-Roe scheme. Both PVM-2U schemes and N-Roe are noticeably less efficient. The differences in efficiency between the analytical and approximated eigenvalue solvers are not significant. For IFCP scheme they seems to be almost identical, but for PVM-2U scheme, the analytical implementation is more efficient.  

\begin{figure}[htbp]
	\center
	\includegraphics[width=6cm]{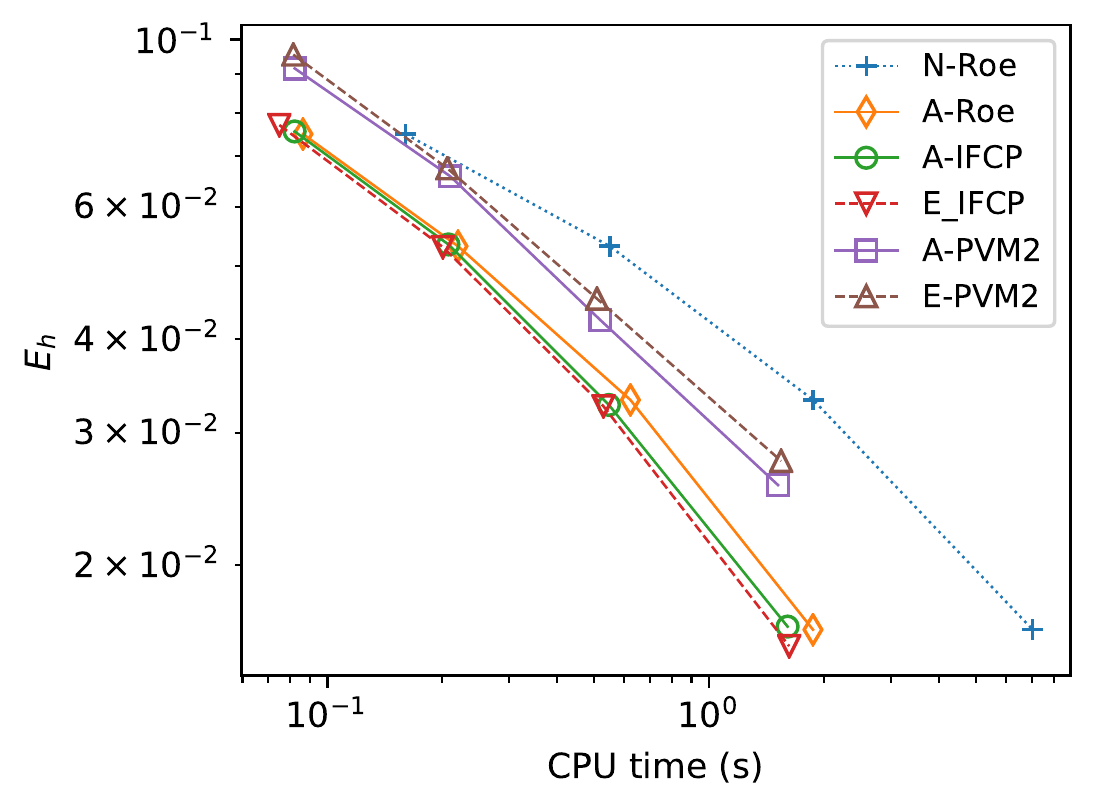}
	\hfill
	\includegraphics[width=6cm]{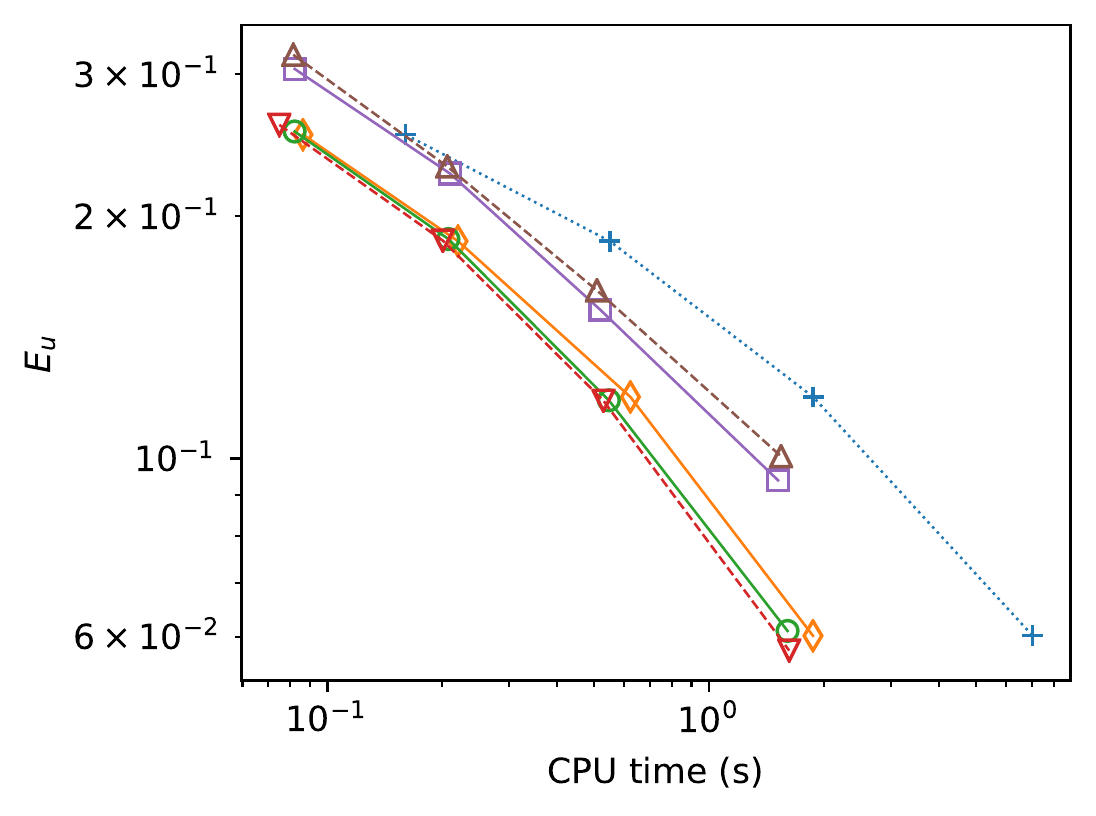}
	\caption{Test V: CPU time vs error for N-Roe, A-Roe, A-IFCP, E-IFCP, A-PVM2, and E-PVM2 scheme, compared to the reference solution}
	\label{fig:Efficiency_Test5}
\end{figure}

\section{Discussion and Conclusion}

This study re-evaluated the efficiency of Roe, IFCP, and PVM-2U schemes for two-layer shallow water systems with different solutions for eigenvalues. For this purpose, numerical, approximated, and recently-proposed analytical solutions for the eigenstructure were considered, combined with Roe, PVM-2U and IFCP numerical schemes.
The choice of eigenvalues in numerical schemes for two-layer SWE models were evaluated in three stages.

First analysis focused only on the accuracy and computational time of different eigenvalue solvers. Numerical and analytical eigenvalue solvers produce almost identical results. Approximated expressions, on the other hand, are less accurate, and the errors grow with density difference between the layers. This increase in errors appears because approximate solutions deviate from the initial assumption that the density ratio is close to one.
Regarding the computational time, analytical and approximated expressions are one order of magnitude faster than the numerical solver, with approximate expressions being two times faster than analytical ones. This additionally confirms our previous study \cite{krvavica2018analytical} where analytical expressions were found to be equally accurate but one order of magnitude faster than numerical solvers (approximated expressions were not considered in the previous study).

The second analysis shifted the focus from eigensolvers to the numerical viscosity matrix, which was computed using different numerical schemes in combination with different eigenvalue solvers. The results revealed that the Roe method is highly sensitive to the choice of eigenvalues, IFCP method is somewhat sensitive, and PVM-2U method shows very little sensitivity to the choice of eigenvalue. 
In general, Roe method is the most accurate, followed by IFCP, and then PVM-2U method. This is expected since IFCP approximates the viscosity matrix using four eigenvalues, and PVM-2U using only two eigenvalues. Also, all schemes are more accurate when analytical eigenvalues are used in comparison to approximated values. It is interesting that the Roe method with approximated eigenvalues is the least accurate method overall for large density differences.
The computational time needed to obtain the numerical viscosity matrix is the longest for Roe scheme with numerical eigensolver. Other combinations of numerical schemes and eigensolvers are several times faster, and very close to each other. The rankings from the fastest to the slowest combination are: PVM-2U scheme with approximated eigenvalues, IFCP with approximated eigenvalues, PVM-2U with analytical eigenvalues, IFCP with analytical eigenvalues, Roe with approximated eigenvalues, and finally Roe with analytical eigenvalues. 
It is important to note that only 10-20\% of total time needed to construct the viscosity matrix is spent on computing the eigenvalues.

In the third analysis several numerical tests were performed to investigate the overall performance of different numerical schemes in combination with different eigenvalue solvers. A total of five tests were designed to evaluate more realistic scenarios, including different density ratios between the layers and various channel geometries. The results revealed that the Roe scheme with approximated eigenvalues is not well-balanced, and should not be considered in two-layer modelling. Remaining four test showed that IFCP (with approximated and analytical eigenvalues) and Roe scheme (with analytical eigenvalues) are very close in performance, with IFCP being slightly better. PVM-2U was noticeably less efficient, regardless of the choice for the eigensolver.

These findings instil more confidence into findings from out previous study \cite{krvavica2018analytical} that showed how Roe method with analytical eigenvalues is very close to IFCP scheme with approximated eigenvalues. Note that the previous study did not considered the modifications to the numerical schemes presented in Section \ref{sec:remarks}, which may improve their computational speed. Also note that the conclusions from the previous study were based only on two numerical examples with the same (small) density difference between the layers and used a fixed time step. Furthermore, the previous study did not evaluate the impact of implementing analytical eigenvalues into PVM and IFCP schemes. Also, the PVM-2U is considered in the present study, whereas \cite{krvavica2018analytical} evaluated PVM-Roe which is a redefinition of the Roe scheme under the PVM paradigm. PVM-2U, on the other hand, is a new PVM method based on two external eigenvalues. 

Overall, when modelling layers of smaller density difference, it seems that approximated eigenvalues are a more efficient choice for IFCP and PVM-2U schemes. For larger density differences, however, the analytical eigenvalues are as efficient. It should be emphasized that analytical eigenvalues are more precise and can help with an accurate prediction of hyperbolicity losses for all density ratios, without producing any overhead in computational time.

The extension of all three considered schemes (Roe, IFCP, PVM-2U) to a higher order is straightforward, following a general approach presented in \cite{castro2006high}. An extension to a two-dimensional case is also possible following the procedure from \cite{castro2009high}. A high order extension of the family of PVM methods is similar to the Roe method \cite{castro2012class}. For example, Castro et al. \cite{castro2012ifcp} presented the extension of IFCP method to solve a two-layer Savage-Hutter type model and simulate tsunamis generated by landslides. Since the eigenvalues affect only the viscosity matrix, the extension of each considered scheme to a higher order is the same regardless of the eigenvalue solver. On the other hand, some impact of the eigenvalues on the efficiency of higher order methods is expected and should be further analysed.

Although the efficiency of analytical solutions to the eigenstructure have been assessed here for two-layer shallow-water flows, these closed-form solutions to eigenvalues expressed in coefficient of a characteristic quartic, can directly be applied to some other non-conservative hyperbolic systems defined by four coupled partial differential equations, such as two-phase granular flows.

\section*{Acknowledgements}

This work has been fully supported by the University of Rijeka under the project number 17.06.2.1.02 (River-Sea Interaction in the Context of Climate Change).

\small
\bibliographystyle{elsarticle-harv} 
\bibliography{Qiqqa2BibTexExport}

\end{document}